\pdfoutput=1
\documentclass[cernpreprint, atlasdraft=false, texlive=2023, UKenglish, texmf, orcidlogo]{atlasdoc}
\usepackage[subfigure=true]{atlaspackage}
\usepackage{atlasbiblatex}

\usepackage{atlasphysics}
\usepackage{multirow}

\usepackage{makecell}

\graphicspath{{logos/}{figures/}}

\usepackage{ANA-HDBS-2021-23-PAPER-defs}

\AtlasTitle{Search for a resonance decaying into a scalar particle and a Higgs boson in final states with leptons and two photons in proton--proton collisions at $\sqrt s=13$~TeV with the ATLAS detector}
\AtlasAbstract{%
A search for a hypothetical heavy scalar particle, $X$, decaying into a singlet scalar particle, $S$, and a Standard Model Higgs boson, $H$, using 140~fb$^{-1}$ of proton--proton collision data at the centre-of-mass energy of 13~TeV recorded with the ATLAS detector at the LHC is presented.
The explored mass range is $300 \leq m_X \leq 1000$~GeV and $170 \leq m_S \leq 500$~GeV.
The signature of this search is one or two leptons ($e$ or $\mu$) from the decay of vector bosons originating from the $S$ particle, $S \rightarrow W^{\pm}W^{\mp}/ZZ$, and two photons from the Higgs boson decay, $H\rightarrow \gamma\gamma$.
No significant excess is observed above the expected Standard Model background. The observed (expected) upper limits at the 95\% confidence level on the cross-section for $gg\rightarrow X\rightarrow SH$, assuming the same $S\rightarrow WW/ZZ$ branching ratios as for a SM-like heavy Higgs boson, are between 530~(800)~fb and 120~(170)~fb.
}

\AtlasRefCode{HDBS-2021-23}

\PreprintIdNumber{CERN-EP-2024-147}

\AtlasJournal{JHEP}
\AtlasJournalRef{JHEP 10 (2024) 104}
\AtlasDOI{10.1007/JHEP10(2024)104}

\hypersetup{pdftitle={ATLAS document},pdfauthor={The ATLAS Collaboration}}

\begin{document}

\maketitle

\tableofcontents

\section{Introduction}
%
%

%

%
The Higgs boson was discovered by the ATLAS~\cite{PERF-2007-01} and CMS~\cite{CMS-CMS-00-001} Collaborations in 2012~\cite{HIGG-2012-27, CMS-HIG-12-028}.
Since then, an important goal has been to determine the Higgs boson properties and to perform precision measurements using proton--proton ($pp$) collision data from the Large Hadron Collider (LHC).
Up until now, all measured properties are consistent with the Standard Model (SM) Higgs boson predictions~\cite{HIGG-2021-23, CMS-HIG-22-001}.
This discovery not only demonstrates the existence of the Higgs boson, but it also opens up new frontiers in particle physics that aim to address the limitations of the SM.
There are a variety of beyond-the-SM scenarios that introduce additional scalar bosons such as the Next-to-Minimal Supersymmetric Standard Model (NMSSM)~\cite{NMSSM1,NMSSM2}, the Two-Real-Singlet Model (TRSM)~\cite{TRSM1,TRSM2} or two-Higgs-Doublet Models (2HDM)~\cite{Branco_2012}.

\begin{figure}[tbhp]
\centering
\includegraphics[width=0.5\textwidth]{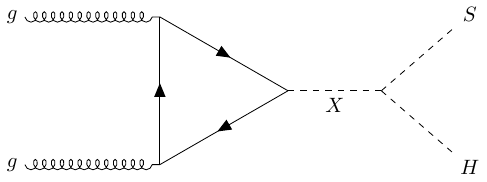}
\caption{Illustrative Feynman diagram for $X \rightarrow SH$ production via gluon--gluon fusion.}
\label{fig:X_SH_FeynmanDiagram}
\end{figure}

The 2HDM+$S$ model~\cite{Baum_2018} extends the 2HDM hypothesis by considering the production of a heavy CP-even scalar boson ($X$) that could decay into an SM Higgs boson ($H$) and a hypothetical scalar singlet ($S$). A representative diagram of the $X\rightarrow SH$ production via gluon--gluon fusion is shown in Figure~\ref{fig:X_SH_FeynmanDiagram}.
The $X\rightarrow SH$ branching ratio is assumed to be 100\%.

Searches inspired by the 2HDM+$S$ probing $X\rightarrow SS\rightarrow WW^{\ast}WW^{\ast}$~\cite{HIGG-2016-24}, $X\rightarrow SH\rightarrow bb\gamma\gamma$~\cite{ATLAS:2024auw} and $X\rightarrow SH\rightarrow VV \tau\tau$~\cite{HDBS-2022-44}, where $V$ can be either a $W^{\pm}$  or $Z$ boson, have been performed by the ATLAS Collaboration.
For the latter, no significant excess is observed above the expected SM processes, and 95\% confidence level (CL) upper limits are set on the signal production cross-section between 542~fb and 72~fb in the mass ranges $500 \leq m_X \leq 1500$~\GeV{} and $200 \leq m_S \leq 500$~\GeV{}~\cite{HDBS-2022-44}.
The $X\rightarrow SH$ search in the $bb\gamma\gamma$ final state~\cite{ATLAS:2024auw} set upper limits on the cross-section times the branching ratio ranging from 39~fb to 0.09~fb, over the mass ranges $170 \leq m_X \leq 1000$~\GeV{} and $15 \leq m_S \leq 500$~\GeV{}.
The CMS Collaboration has also performed searches for $X\rightarrow SH$ in the $bb\gamma\gamma$~\cite{CMS-HIG-21-011}, $bb\tau\tau$~\cite{CMS-HIG-20-014}, and $4b$~\cite{CMS-B2G-21-003} decay modes.
In the diphoton plus two $b$-quarks search, the upper limits on the product of the production cross-section and the decay branching ratios of the signal process lie in the range of $0.9 - 0.04$~fb~\cite{CMS:2023boe}, in the explored mass ranges $300 \leq m_X  \leq 1000$~\GeV{} and $90 \leq m_S \leq 800$~\GeV{}.
For the search with a pair of tau leptons and $b$-quarks in the final state, limits are set on the production cross-section ranging from 125~fb to 2.7~fb in the mass ranges  $240 \leq m_X \leq 3000$~\GeV{} and $60 \leq m_S \leq 2800$~\GeV{}~\cite{CMS-HIG-20-014}.
Comparable limits are found in the search involving four $b$-quarks in the final state, ranging from 150 to 0.1~fb in the mass ranges $0.9 \leq m_X \leq  4$~\TeV{} and $60 \leq m_S \leq 600$~\GeV{}~\cite{CMS-B2G-21-003}.
All these results are obtained assuming for the $S$ boson the same mass-dependent branching ratios as for the SM Higgs boson~\cite{deFlorian:2016spz} (denoted SM-like branching ratios in the following).

This paper is focused on the search for $X\rightarrow SH \rightarrow VV\gamma\gamma$.
The final state of interest is characterised by two photons from the SM Higgs boson decay ($H\rightarrow \gamma\gamma$), and one or two leptons (electrons or muons) originating from the vector bosons produced in the $S\rightarrow VV$ decays.
This signature benefits from a good diphoton mass ($m_{\gamma\gamma}$) resolution~\cite{HIGG-2019-13} and the $m_{\gamma\gamma}$ distribution is used as the final discriminant.
The requirement of at least one lepton rejects some SM background processes and therefore increases the signal-to-background ratio.
The events are classified by the number and flavour of the leptons (electrons and muons) in the final state and multivariate analysis techniques are used to enhance the sensitivity of the search.
The algorithm is trained to distinguish between the dominant SM backgrounds (multi-jet processes and vector bosons produced in association with a pair of photons) and the $X\rightarrow SH\rightarrow VV\gamma\gamma$ signal.
This search is performed over the mass ranges $300 \leq m_X \leq 1000$~\GeV{} and  $170 \leq m_S \leq 500$~\GeV{}.
It allows to explore lower mass ranges than other final states with $b$-quarks where the signal becomes boosted at low $m_S$ values, and suffer from low energetic $b$-quarks falling below the reconstructed threshold in the low $m_X$ region.
In the interpretation of the search results, the $S$ boson is assumed to have SM-like branching ratios.
Two additional scenarios wherein the $S$ boson decays with a 100\% branching ratio into a pair of $W^{\pm}$ or $Z$ bosons, $S\rightarrow W^{+}W^{-}/ZZ$, are also considered.

This paper is organised as follows.
A brief description of the ATLAS detector is given in Section~\ref{sec:Detector}.
Data and simulation samples are described in Section~\ref{sec:Samples}.
The object reconstruction and event selection are outlined in Section~\ref{sec:Analysis}.
The background estimation and the systematic uncertainties are described in Section~\ref{sec:BkgEstimation} and Section~\ref{sec:Systematic}, respectively.
Section~\ref{sec:Result} presents the results of this search, which are summarised in Section~\ref{sec:Conclusion}.
%


\FloatBarrier
\section{ATLAS detector}
\label{sec:Detector}
%

%
%
%
%
%
%
%
%
%

\newcommand{\AtlasCoordFootnote}{%
ATLAS uses a right-handed coordinate system with its origin at the nominal interaction point (IP)
in the centre of the detector and the \(z\)-axis along the beam pipe.
The \(x\)-axis points from the IP to the centre of the LHC ring,
and the \(y\)-axis points upwards.
Polar coordinates \((r,\phi)\) are used in the transverse plane,
\(\phi\) being the azimuthal angle around the \(z\)-axis.
The pseudorapidity is defined in terms of the polar angle \(\theta\) as \(\eta = -\ln \tan(\theta/2)\) and is equal to the rapidity
$ y = \frac{1}{2} \ln \left( \frac{E + p_z c}{E - p_z c} \right) $ in the relativistic limit.
Angular distance is measured in units of \(\Delta R \equiv \sqrt{(\Delta y)^{2} + (\Delta\phi)^{2}}\).}

The ATLAS detector~\cite{PERF-2007-01} at the LHC covers nearly the entire solid angle around the collision point.\footnote{\AtlasCoordFootnote}
It consists of an inner tracking detector surrounded by a thin superconducting solenoid, electromagnetic and hadronic calorimeters,
and a muon spectrometer incorporating three large superconducting air-core toroidal magnets.

The inner-detector system (ID) is immersed in a \qty{2}{\tesla} axial magnetic field
and provides charged-particle tracking in the range \(|\eta| < 2.5\).
The high-granularity silicon pixel detector covers the vertex region and typically provides four measurements per track,
the first hit generally being in the insertable B-layer (IBL) installed before Run~2~\cite{ATLAS-TDR-19,PIX-2018-001}.
It is followed by the SemiConductor Tracker (SCT), which usually provides eight measurements per track.
These silicon detectors are complemented by the transition radiation tracker (TRT),
which enables radially extended track reconstruction up to \(|\eta| = 2.0\).
The TRT also provides electron identification information
based on the fraction of hits (typically 30 in total) above a higher energy-deposit threshold corresponding to transition radiation.

The calorimeter system covers the pseudorapidity range \(|\eta| < 4.9\).
Within the region \(|\eta|< 3.2\), electromagnetic calorimetry is provided by barrel and
endcap high-granularity lead/liquid-argon (LAr) calorimeters,
with an additional thin LAr presampler covering \(|\eta| < 1.8\)
to correct for energy loss in material upstream of the calorimeters.
Hadronic calorimetry is provided by the steel/scintillator-tile calorimeter,
segmented into three barrel structures within \(|\eta| < 1.7\), and two copper/LAr hadronic endcap calorimeters.
The solid angle coverage is completed with forward copper/LAr and tungsten/LAr calorimeter modules
optimised for electromagnetic and hadronic energy measurements respectively.

The muon spectrometer (MS) comprises separate trigger and
high-precision tracking chambers measuring the deflection of muons in a magnetic field generated by the superconducting air-core toroidal magnets.
The field integral of the toroids ranges between \num{2.0} and \qty{6.0}{\tesla\metre}
across most of the detector.
Three layers of precision chambers, each consisting of layers of monitored drift tubes, cover the region \(|\eta| < 2.7\),
complemented by cathode-strip chambers in the forward region, where the background is highest.
The muon trigger system covers the range \(|\eta| < 2.4\) with resistive-plate chambers in the barrel, and thin-gap chambers in the endcap regions.

The luminosity is measured mainly by the LUCID--2~\cite{LUCID2} detector that records Cherenkov light produced in the quartz windows of photomultipliers located close to the beampipe.

Events are selected by the first-level trigger system implemented in custom hardware,
followed by selections made by algorithms implemented in software in the high-level trigger~\cite{TRIG-2016-01}.
The first-level trigger accepts events from the \qty{40}{\MHz} bunch crossings at a rate below \qty{100}{\kHz},
which the high-level trigger further reduces in order to record complete events to disk at about \qty{1}{\kHz}.

A software suite~\cite{SOFT-2022-02} is used in data simulation, in the reconstruction
and analysis of real and simulated data, in detector operations, and in the trigger and data acquisition
systems of the experiment.


\FloatBarrier
\section{Data and simulation samples}
\label{sec:Samples}

\subsection{Data Samples}
The data used were collected with the ATLAS detector during 2015--2018, from $pp$ collisions at a centre-of-mass energy $\sqrt{s}=13$~\TeV{},
corresponding to an integrated luminosity of 140~\ifb{} with an uncertainty of 0.83\%~\cite{DAPR-2021-01} after data quality requirements~\cite{DAPR-2018-01}.
Events were recorded using diphoton triggers that require two reconstructed photon candidates with minimum transverse energies of 35~\GeV{} and 25~\GeV{}~\cite{TRIG-2018-05}.
During the 2015--2016 data taking period, a \textit{Loose} identification requirement was applied for this diphoton trigger while it was replaced by the \textit{Medium} selection criteria to keep a tolerable trigger rate in 2017--2018 due to the increased instantaneous luminosity.

\subsection{Monte Carlo simulated samples}

\subsubsection{Signal samples}
The Monte Carlo (MC) simulated signal samples were produced with the \PYTHIA[8] generator~\cite{Sjostrand:2014zea} with the matrix element calculation at leading order (LO) accuracy in quantum chromodynamics (QCD), followed by
parton showering, hadronisation and underlying event modelling using the A14 set of tuned parameters (``tune'')~\cite{ATL-PHYS-PUB-2014-021} and the \NNPDF[2.3lo] parton distribution functions (PDF)~\cite{Ball:2012cx}.
During the sample generation, both $X$ and $S$ were assumed to have a narrow width compared with the experimental resolution, and their widths are fixed to 10~\MeV{}.
A total of 20 signal samples for various $m_{X}$ and $m_{S}$ were generated. %
The $X$ boson was required to decay into $S$ and $H$ with $S$ only decaying into a pair of $W$ or $Z$ bosons and $H$ decaying into a pair of photons.
By considering leptonic decays of $W$ or $Z$ bosons, the following three final state samples were produced for each $m_{X}$ and $m_{S}$ combination: \gglvqq{}, \gglvlv{}, and \ggllqq{}, where $\ell=e$, $\mu$, or $\tau$.
The \ggllll{} decay sample is excluded due to its negligible contribution.
To achieve a better signal generation efficiency, the samples were produced by requiring to have at least one lepton with transverse momentum ($\pt$) greater than 7~\GeV{} and pseudorapidity $|\eta|<3$ at the generator level.

\subsubsection{Background samples}
\label{sec:bkg_sample}

The main background contributions result from SM single and double-Higgs boson production, forming a resonance on the diphoton mass ($m_{\gamma\gamma}$) spectrum, and other SM processes giving a smoothly falling $m_{\gamma\gamma}$ spectrum (continuum background).
The corresponding events were generated with MC simulation.

Simulated events for single Higgs boson production via gluon--gluon fusion (ggF) were produced with the \POWHEGBOX[v2] generator~\cite{Hamilton:2013fea,Hamilton:2015nsa,Alioli:2010xd,Nason:2004rx,Frixione:2007vw} at next-to-next-to-leading order (NNLO) accuracy in QCD and interfaced with \PYTHIA[8].
The NNLO accuracy for arbitrary inclusive $gg\to H$ observables was achieved by reweighting the Higgs boson rapidity spectrum in \textsc{Hj}-\MINLO~\cite{Hamilton:2012np,Campbell:2012am,Hamilton:2012rf} to that of HNNLO~\cite{Catani:2007vq}.
The \PDFforLHC[15nnlo] PDF set~\cite{Butterworth:2015oua} and the \AZNLO tune~\cite{STDM-2012-23} of \PYTHIA[8] were used and the decays of $b$- and $c$-hadrons were modelled by the \EVTGEN[1.6.0] programme~\cite{Lange:2001uf}.
These events were normalised using the NNLO cross-section in QCD plus electroweak corrections
at next-to-leading order (NLO)~\cite{deFlorian:2016spz,Anastasiou:2016cez,Anastasiou:2015ema,Dulat:2018rbf,Harlander:2009mq,Harlander:2009bw,Harlander:2009my,Pak:2009dg,Actis:2008ug,Actis:2008ts,Bonetti:2018ukf,Bonetti:2018ukf}.

Simulated single Higgs boson events produced via vector-boson fusion (VBF) were generated with \POWHEGBOX[v2] at NLO accuracy in QCD and interfaced with \PYTHIA[8].
The \PDFforLHC[15nlo] PDF set and \AZNLO tune were used.
Simulated events were normalised using an approximate-NNLO QCD cross-section with NLO electroweak corrections~\cite{Ciccolini:2007jr,Ciccolini:2007ec,Bolzoni:2010xr}.

Events of single Higgs boson produced in association with a vector boson ($VH$, $V=W/Z$) were simulated using \POWHEGBOX[v2] and interfaced with \PYTHIA[8].
The \POWHEG prediction is accurate to NLO in QCD for $VH+$1jet distributions by using the \MINLO~\cite{Luisoni:2013cuh} prescription.
The loop-induced $gg\to ZH$ process was generated separately at LO.
The \PDFforLHC[15nlo] PDF set and the \AZNLO tune were used.
Cross-sections calculated at NNLO in QCD with NLO electroweak corrections for $q\bar{q}/qg \to VH$ and at NLO and next-to-leading-logarithm accuracy in QCD for $gg \to ZH$~\cite{Ciccolini:2003jy,Brein:2003wg,Brein:2011vx,Altenkamp:2012sx,Denner:2014cla,Brein:2012ne,Harlander:2014wda} were used for the normalisation of the MC samples.

Events corresponding to Higgs boson production in association with a pair of top or bottom quarks ($t\bar{t}H$ or $b\bar{b}H$) were simulated using \POWHEGBOX[v2] at NLO with the \NNPDF[3.0nlo] PDF set~\cite{Ball:2014uwa}.
The events were interfaced with \PYTHIA[8] using the A14 tune and the \NNPDF[2.3lo] PDF set.
The decays of bottom and charm hadrons were performed with \EVTGEN[1.6.0].

Finally, events for single Higgs boson production in association with a single top quark were simulated with the \textsc{MadGraph}\_ \textsc{aMC@NLO}~2.3.3~\cite{Alwall:2014hca} generator at NLO with the \NNPDF[3.0nlo] PDF set.
The events were interfaced with \PYTHIA[8] using the A14 tune and the \NNPDF[2.3lo] PDF set.

In addition to the single Higgs boson processes, events corresponding to the SM double Higgs boson ggF and VBF production modes are also considered.
Those events were generated with \POWHEGBOX[v2] at NLO accuracy in QCD using the \PDFforLHC[15nlo] PDF set and interfaced with \PYTHIA[8].
During the sample generation, one of the Higgs bosons was required to decay into two photons and the other Higgs boson was required to decay into $WW$, $ZZ$ or $\tau\tau$, giving a final state with a pair of electrons or muons, or with one electron and one muon.
Leptons were required to have $\pt>7\,\GeV$ and $|\eta|<3$ at the generator level.

The normalisation of all Higgs boson samples accounts for the decay branching ratios calculated with HDECAY~\cite{Djouadi:1997yw,Spira:1997dgv2,Djouadi:2006bz}
and \PROPHECY~\cite{Bredenstein:2006ha,Bredenstein:2006rh,Bredenstein:2006nkv2}.

Continuum background from $\gamma\gamma$+jets, $V$+$\gamma\gamma$, and $t\bar{t}$+$\gamma\gamma$ processes is considered.
Their contributions are described with corresponding MC simulated samples that are exclusively used for the event selection optimisation.
These samples were normalised with cross-sections as predicted by their corresponding MC generators.

Events from $\gamma\gamma$+jets production were simulated using the \SHERPA[2.2.4] generator~\cite{Sherpa:2019gpd} at NLO accuracy in QCD with up to one additional parton and at LO with up to three additional partons.
The matrix elements of these events were calculated with the Comix~\cite{Comix} and \OPENLOOPS~\cite{OpenLoop1,OpenLoop2} libraries and then matched to the \SHERPA parton shower~\cite{SherpaShower} using the MEPS@NLO prescription~\cite{MEPS3,MEPS4,MEPS1,MEPS2}.
The \NNPDF[3.0nnlo] PDF set~\cite{Ball:2014uwa} was used to describe the parton distributions in the incoming protons.
A generator-level selection was applied to these events with the requirement of the invariant mass of the two photons to be between 90~\GeV{} and 175~\GeV{}.

The $V+\gamma\gamma$ events were generated using \SHERPA[2.2.4] at NLO accuracy in QCD with up to one additional parton and up to three extra partons at LO.
The calculation procedure is the same as in $\gamma\gamma$+jets event generation.
Events were generated separately according to their final states as listed: $ee$+$\gamma\gamma$, $\mu\mu$+$\gamma\gamma$, $\tau\tau$+$\gamma\gamma$, $e\nu$+$\gamma\gamma$, $\mu\nu$+$\gamma\gamma$, $\tau\nu$+$\gamma\gamma$, and $\nu\nu$+$\gamma\gamma$.
The generator-level photon \pt{} was required to be greater than 17~\GeV{} and the invariant mass of the two photons should be larger than 80~\GeV{} for these events.

The $t\bar{t}$+$\gamma\gamma$ process is simulated with the \MGNLO generator at LO and interfaced with \PYTHIA[8].
The \NNPDF[2.3lo] PDF set and A14 tune were used for this production.
The decays of bottom and charm hadrons were performed with \EVTGEN[1.6.0].

In addition, dedicated samples (denoted ``lepton-dependence samples'') corresponding to final states of $\gamma\gamma+0\ell+\mathrm{jets}$, $\gamma\gamma+1 \ell+\mathrm{jets}$, and $\gamma\gamma+2\ell+\mathrm{jets}$ are generated to study the $\myy$ distribution difference for cases with different lepton multiplicity at the generator level. %
These samples were produced with \MGNLO{} interfaced with \PYTHIA[8].
All possible SM processes with described final states except for those with $H\to\gamma\gamma$ were included in the event generation.

All simulated events except for the signals, $\gamma\gamma$+jets and lepton-dependence samples were passed through a detailed detector simulation of the ATLAS detector implemented with \GEANT~\cite{Agostinelli:2002hh,ATLASFullSim}.
The remaining samples were simulated using $\textsc{AtlFastII}$~\cite{ATLASFullSim}, which exploys \GEANT{} except for a parameterisation of the calorimeter response~\cite{ATL-PHYS-PUB-2010-013}.
The effect of multiple interactions in the same and neighbouring bunch crossings (\pileup) was modelled by overlaying the simulated hard-scattering event with inelastic $pp$ events generated with \PYTHIA[8]~\cite{Sjostrand:2007gs}
using the \NNPDF[2.3lo] PDF set and the A3 tune~\cite{ATL-PHYS-PUB-2016-017}.
A summary of MC simulated samples can be found in Table~\ref{tab:summary_sample}.

%
%
%
%
%
%
%
%
%
%
%
%
%
%
%
%
%
%
%
%
%
%
%
%
%
%
%
%
%
%
%
%
%
%
%
%
%
%
%
%
%
%
%
%
%
%
%
%
%
%
%
%
%
%
%
%
%
%
%
%
%
%
%
%
%
%
%
%
%
%
%
%
%
%
%
%
%
%
%
%
%
%
%
%
%
%

\begin{table}[t]
\renewcommand{\arraystretch}{1.1}
\begin{center}
\caption{Summary of MC simulated samples used in this analysis.}
\label{tab:summary_sample}
\begin{tabular}{c|c|c|c}
\hline
\hline
Process                                                 & Generator                          & PDF                   & Tune         \\
\hline
\multicolumn{4}{c}{Signal}                                                                                                                       \\
\hline
$X\rightarrow SH\rightarrow VV+\gamma\gamma $ & \textsc{Pythia\,8}                            & NNPDF2.3\textsc{lo}  &            A14        \\
\hline
\multicolumn{4}{c}{SM Single and double Higgs boson production}                                                                                 \\
\hline
ggF $H$                                                             & \textsc{Powheg}+\textsc{Pythia\,8}            & PDF4LHC15\textsc{nnlo}         &      AZNLO    \\
VBF $H$                                                            & \textsc{Powheg}+\textsc{Pythia\,8}            & PDF4LHC15\textsc{nlo}          &     AZNLO     \\
$WH$                                                             & \textsc{Powheg}+\textsc{Pythia\,8}            & PDF4LHC15\textsc{nlo}          &     AZNLO     \\
$qq\to ZH$                                                       & \textsc{Powheg}+\textsc{Pythia\,8}            & PDF4LHC15\textsc{nlo}          &     AZNLO     \\
$gg\to ZH$                                                       & \textsc{Powheg}+\textsc{Pythia\,8}            & PDF4LHC15\textsc{nlo}          &     AZNLO     \\
$t\bar{t}H$                                                            & \textsc{Powheg}+\textsc{Pythia\,8}            & NNPDF3.0\textsc{nlo} & A14 \\
$b\bar{b}H$                                                            & \textsc{Powheg}+\textsc{Pythia\,8}            & NNPDF3.0\textsc{nlo} & A14 \\
$tHbj$                                                           & \textsc{MadGraph\_aMC@NLO}+\textsc{Pythia\,8} & NNPDF3.0\textsc{nlo} & A14 \\
$tHW$                                                            & \textsc{MadGraph\_aMC@NLO}+\textsc{Pythia\,8} & NNPDF3.0\textsc{nlo} & A14 \\
ggF $HH\to VV+ \gamma\gamma$                   & \textsc{Powheg}+\textsc{Pythia\,8}            & PDF4LHC15\textsc{nlo}          &  A14        \\
VBF $HH\to VV+ \gamma\gamma$                   & \textsc{Powheg}+\textsc{Pythia\,8}            & PDF4LHC15\textsc{nlo}          &  A14        \\
\hline
\multicolumn{4}{c}{Continuum background}                                                                                                         \\
\hline
$\gamma\gamma+\mathrm{jets}$                                     & \textsc{Sherpa}                             & NNPDF3.0\textsc{nnlo}            &  --      \\
$V+\gamma\gamma$                                              & \textsc{Sherpa}                             & NNPDF3.0\textsc{nnlo}               &  --   \\
$t\bar{t}\gamma\gamma$                                           & \textsc{MadGraph\_aMC@NLO}+\textsc{Pythia\,8} & NNPDF3.0\textsc{nlo}  & A14 \\
\hline
\multicolumn{4}{c}{Lepton-dependence samples}                                                                                                    \\
\hline
$\gamma\gamma+0\ell+\mathrm{jets}$                                             & \textsc{MadGraph\_aMC@NLO}+\textsc{Pythia\,8} & NNPDF3.0\textsc{nlo}  & A14 \\
$\gamma\gamma+1\ell+\mathrm{jets}$                                        & \textsc{MadGraph\_aMC@NLO}+\textsc{Pythia\,8} & NNPDF3.0\textsc{nlo} & A14  \\
$\gamma\gamma+2\ell+\mathrm{jets}$                                    & \textsc{MadGraph\_aMC@NLO}+\textsc{Pythia\,8} & NNPDF3.0\textsc{nlo} & A14  \\
\hline
\hline
\end{tabular}
\end{center}
\end{table}



%

%
%
%
%
%

\FloatBarrier
\section{Object and event selection}
\label{sec:Analysis}

\subsection{Object selection}
\label{sec:ObjectSelection}

%
%

Vertices from $pp$ collisions are reconstructed if they have associated at least two ID tracks with $\pT > 0.5$~\GeV{}.
The diphoton primary vertex (PV) is chosen by using a neural network algorithm that uses information about the ID tracks as well as the photon candidates~\cite{HIGG-2013-08}.

Photons are reconstructed based on a dynamic, topological cell clustering-based algorithm from the energy deposits in the electromagnetic calorimeter in the region $|\eta| < 2.37$, excluding the transition region between the barrel and endcap calorimeters $1.37 < |\eta| < 1.52$~\cite{EGAM-2018-01}.
The photon identification criteria is constructed using information from the shower shapes and the primary identification criteria is labelled as $Tight$.
The photon isolation criteria quantifies the activity near the photons from the tracks of nearby charged particles, or from energy deposits in the calorimeter~\cite{EGAM-2018-01}.
This analysis considers events by selecting photon candidates which are required to satisfy a set of preselection criteria.
The two photons with the highest transverse momentum, referred to as leading ($\gamma_1$) and subleading ($\gamma_2$) photons, must satisfy $\pT > 22$~\GeV{} and  $|\eta| < 2.37$, excluding the transition region between the barrel and endcap calorimeters $1.37 < |\eta| < 1.52$.
Photon candidates are separated from multi-jet backgrounds by applying $Tight$ identification and further isolation requirements to suppress jets misidentified as photons.

Electrons are reconstructed and identified based on clusters built from energy deposits in the electromagnetic calorimeter, which are matched to a track in the inner detector~\cite{EGAM-2018-01}.
The muon reconstruction is performed using information from the inner detector and muon spectrometer, as well as the electromagnetic and hadronic calorimeters~\cite{MUON-2018-03}.
In this search, electron candidates are required to have $\pT > 10$~\GeV{} and $|\eta|<2.47$, excluding the transition region between the barrel and endcap calorimeters $1.37 < |\eta| < 1.52$.
Muon candidates should have $\pT > 10$~\GeV{} and $|\eta|<2.7$.
Leptons must satisfy $Medium$ identification and $Loose$ isolation~\cite{EGAM-2018-01,MUON-2018-03}, and a set of requirements based on the longitudinal and transverse impact parameters relative to the vertex and the beam axis.

Jets are reconstructed using a particle flow algorithm~\cite{PERF-2015-09} from noise-suppressed positive-energy topological clusters~\cite{PERF-2014-07} in the calorimeter using the anti-$k_t$ algorithm~\cite{FastJet,antiKt} with a radius parameter $R=0.4$.
The jet energy scale calibration restores the jet $\pT$, energy, and mass to that of jets reconstructed at particle level~\cite{JETM-2018-05}.
In this search, jets are required to have $\pT > 25$~\GeV{} and to be in the central region of the detector, $|\eta|<2.5$.
To suppress jets from pileup, a jet-vertex-tagger multivariate discriminant~\cite{PERF-2014-03} is applied to jets with $\pT < 60$~\GeV{}.
Jets containing $b$-hadrons are identified ($b$-tagged) using the 77\% efficiency working point of the DL1r $b$-tagging algorithm~\cite{FTAG-2019-07}.

An overlap removal procedure is performed to avoid double-counting objects.
First, electrons overlapping with any of the two selected photons ($\Delta R < 0.4$) are removed.
Jets overlapping with the selected photons ($\Delta R < 0.4$) or electrons ($\Delta R < 0.2$) are removed.
Electrons overlapping with the remaining jets ($\Delta R < 0.4$) are removed to match the requirements imposed when measuring isolated electron efficiencies.
Finally, muons overlapping with photons or jets ($\Delta R < 0.4$) are removed.

The missing transverse momentum, with magnitude \MET, is defined as the negative vector sum of the transverse momenta of the selected photon, electron, muon, and jet objects,
as well as of the transverse momenta of remaining low-$\pT$
particles estimated by using tracks associated with the diphoton primary vertex but not assigned to any of the selected objects~\cite{JETM-2020-03}.

The above requirements constitute the event preselection of this search.


\subsection{Event selection}
This search selects events with two photons from the Higgs boson decay, and one or two leptons coming from the vector bosons originated from the $S\rightarrow VV$ process.
Events are required to pass diphoton triggers as described in Section~\ref{sec:Samples}. %
Moreover, photons are required to have $\pT^{\gamma_{1(2)}}/m_{\gamma\gamma} >$ 0.35 (0.25)~\cite{HIGG-2019-13}.
Selected events must contain one or two additional leptons ($e$ or $\mu$) with $\pT > 10$~\GeV{}.
To suppress backgrounds with top quarks, events containing one or more $b$-tagged jets are rejected.

The events are classified into four different regions depending on the number and flavour of leptons originating from the vector boson decays in the $S\rightarrow W^{\pm}W^{\mp}$ and  $S\rightarrow ZZ$ processes.
Events in the one-lepton ($1\ell$) region are required to have one lepton and at least two jets.
The other regions account for events with two leptons with opposite electric charge. Events with two leptons of different flavour are targeted in the $e\mu$ region.
Events having two leptons of same flavour (SF) are further split by checking the compatibility of the dilepton invariant mass with the $Z$-mass pole.
Events with at least two jets and satisfying $|m_{\ell\ell} - 91.2\,\GeV| < 10$~\GeV{} are classified in the $2\ell(ZZ)$ region, which targets the $S\rightarrow ZZ\rightarrow \ell\ell$+jets process.
The remaining SF events are included in the $2\ell(WW)$ region.

Two optimisation strategies are adopted.
For each of the $1\ell$ and $2\ell(WW)$ regions, a Boosted Decision Tree (BDT) is used to enhance the analysis sensitivity.
The $1\ell$ and $2\ell(WW)$ regions further split the events by dividing the BDT output distribution into the loose (low BDT score) and tight (high BDT score) signal regions.
The $e\mu$ and $2\ell(ZZ)$ signal regions are limited in statistics and have higher signal-over-background ratios than the BDT-based regions.
Due to this, the $e\mu$ and $2\ell(ZZ)$ signal regions follow an inclusive cut-based strategy and events are not further split into sub-regions.
This analysis strategy results in six signal regions.

The final discriminant of this search is the diphoton invariant mass spectrum. %
To be consistent with the $H\rightarrow \gamma\gamma$ process and to exclude the region of the $Z$-boson resonance, the range of $m_{\gamma\gamma}$ is limited to the $[105,160]$~\GeV{} region.
The signal region is defined as the $m_{\gamma\gamma}$ within $[120,~130]$~\GeV{}.
Events outside the signal region, referred to as sideband events, are used to estimate the main background processes as described in Section~\ref{sec:BkgEstimation}.
After all selections described above are applied, the combined acceptance and selection efficiency for the signal production ranges from 11\% to 25\%, which increases for higher $m_X$ hypotheses. %
Table~\ref{tab:selection} summarises the event selection and strategy for each of the signal regions.

%

\begin{table}[t]
\renewcommand{\arraystretch}{1.1}
\begin{center}
\caption{Event selection and classification strategy.}%
\label{tab:selection}

\begin{tabular}{l | c c c c c c }
\hline\hline
Preselection                     & \multicolumn{6}{c}{Two photon candidates and no $b$-tagged jets}                                         \\
\hline
Region                          &  \multicolumn{2}{c}{$1\ell$}                                                       & $e\mu$    & \multicolumn{2}{c}{$2\ell(WW)$} & $2\ell(ZZ)$ \\
\hline
Number of leptons                 & \multicolumn{2}{c}{1}                                                              & 2         & \multicolumn{2}{c}{2}           & 2           \\
Total electric charge             & \multicolumn{2}{c}{--}                                                              & 0         & \multicolumn{2}{c}{0}           & 0           \\
Same flavour leptons              & \multicolumn{2}{c}{--}                                                              & No        & \multicolumn{2}{c}{Yes}         & Yes         \\
$|m_{\ell\ell} - m_{Z}|$~[\GeV{}]    & \multicolumn{2}{c}{--}                                                              & --         & \multicolumn{2}{c}{$>10$}       & $<10$       \\
Number of jets                    & \multicolumn{2}{c}{$\geq 2$}                                                       & --         & \multicolumn{2}{c}{-}           & $\geq 2$    \\
\hline
Strategy                          & \multicolumn{2}{c}{BDT}                                                            & Cut-based & \multicolumn{2}{c}{BDT}         & Cut-based   \\
Number of signal regions &  \multicolumn{2}{c}{2} & 1 & \multicolumn{2}{c}{2} & 1 \\
\hline
$m_{\gamma\gamma}$ region      & \multicolumn{6}{c}{$[105,160]$~\GeV{}}                                              \\
\hline
\hline
\end{tabular}

\end{center}
\end{table}


%
%
%
%
%
%
%
%
%
%
%
%
%

%
%
%

%
%
\subsection{Boosted Decision Tree strategy}
The dominant signal process in this search is the $S\rightarrow W^{+}W^{-}$ decay given the larger branching ratio compared with the $S\rightarrow ZZ$ process for the explored $m_S$ range.
In the $1\ell$ region one of the $W$ bosons decays leptonically and the other decays hadronically.
In the $e\mu$ and $2\ell(WW)$ regions both $W$ bosons decay leptonically resulting into a pair of leptons with different and same flavour, respectively.
Two BDTs are built based on the kinematic observables from the final-state objects in the $1\ell$ and $2\ell(WW)$ regions.
The different signal samples are grouped according to the $S$ mass into four groups: $m_S=170$~\GeV{}, $m_S=200$~\GeV{}, $m_S=300$~\GeV{} and $m_S=400,500$~\GeV{}.
These groups contain from four to six signal samples depending on the $m_X$ values in each group.
The BDT algorithms are trained for each group against the total background from MC simulation using the parameterised BDT method~\cite{Baldi2016}.

Twelve and nine variables for the $1\ell$ and $2\ell(WW)$ regions respectively are used to train each BDT, as listed in Table~\ref{tab:MVA_Variables}.
The BDT input variables list excludes $m_{\gamma\gamma}$ as it is used as the main discriminant; these variables are selected to have small correlation with $m_{\gamma\gamma}$.   %
The BDT variable with the highest separation power is the transverse momentum of the pairs of photons from the SM Higgs boson decay ($\pt^{\gamma\gamma}$).
The comparison of the  $\pt^{\gamma\gamma}$ distributions for data, the expected SM background processes and the $(m_X, \,m_S) = (1000,\, 300)$~\GeV{} signal from simulation is shown in Figure~\ref{fig:pTyy} for the $1\ell$ and $2\ell(WW)$ regions.

\begin{table}[t]
\renewcommand{\arraystretch}{1.1}
\begin{center}
\caption{Variables used as inputs to the BDT in the $1\ell$ and $2\ell(WW)$ regions.  The highest-\pt\ (leading) lepton is denoted $\ell_1$, and the subleading lepton is denoted $\ell_2$. The numbers indicate the ranking of each input variable, with 1 corresponding to the most highly ranked variable.}
\resizebox{\textwidth}{!}{
\begin{tabular}{ l l c c }
\hline
\hline
\multirow{2}{*}{Variable}      & \multirow{2}{*}{Description}                              & \multicolumn{2}{c}{BDT-based regions}                              \\
&                                                                                         & $1\ell$ & $2\ell(WW)$  \\ \toprule
$\Delta R(\gamma\gamma,\ell\nu jj)$     & \dR\ between the diphoton system and the $\ell + \met jj$ system   &                                 10          &                  \\
$\Delta R(\gamma\gamma,\ell\nu\ell\nu)$ & \dR\ between the diphoton system and the $\ell\ell + \met$ system  &                                           &          9                   \\ \midrule
$\Delta R(jj,\ell\nu)$                  & \dR\ between the dijet system and the $\ell + \met$ system         &                                     9      &                  \\
$\Delta R(\ell_1\nu,\ell_2)$            & \dR\ between leading lepton + \met and subleading lepton           &                                           &       8                      \\
$\pt^{\ell + \met jj}$                   & \pt\ of the $\ell + \met jj$ system                                &                                    2       &                   \\ \midrule
$\pt^{\gamma\gamma}$  & \pt\ of the diphoton system                                        & 1                                          & 2                  \\
$\Delta\phi(\yy, \ell_{(1)})$           & $\Delta \phi$ between the diphoton system and the (leading) lepton & 12                                         & 7                  \\ \midrule
$\Delta R(\ell,\met)$                   & \dR\ between the lepton and the \met                               &                     8                      &                   \\
$\pt^{\ell_{(1)}}$                       & \pt\ of the (leading) lepton                                       & 4                                          &  4                  \\
$\pt^{\ell_{1}+\met}$                    & \pt\ of the leading lepton and \met system                         &                                           &        3                     \\
$m_{\textrm{T}}(\ell_{(1)} \met)$         & Transverse mass of the (leading) lepton and \met               & 11                                          & 5                  \\
$m_{\ell\ell}$                           & Invariant mass of the dilepton system                          &                                           &                  6           \\
\met                                    & Missing transverse energy                                          & 3                                          & 1                  \\ \midrule
$\Delta R(j,j)$                         & \dR\ between the two jets  with closest mass to $m_W$                           &                       6                    &                   \\
$\pt^{jj}$                               & \pt\ of the the two jets  with closest mass to $m_W$                                   &                          5                 &                  \\
$m_{jj}$                                 & Invariant mass of the dijet system with closest mass to $m_W$     &                 7                          &                   \\  %
\hline
\hline
\end{tabular}
}
\label{tab:MVA_Variables}
\end{center}
\end{table}

\begin{figure}[tbhp]
\begin{center}
\subfigure[]{\includegraphics[width = 0.4\textwidth]{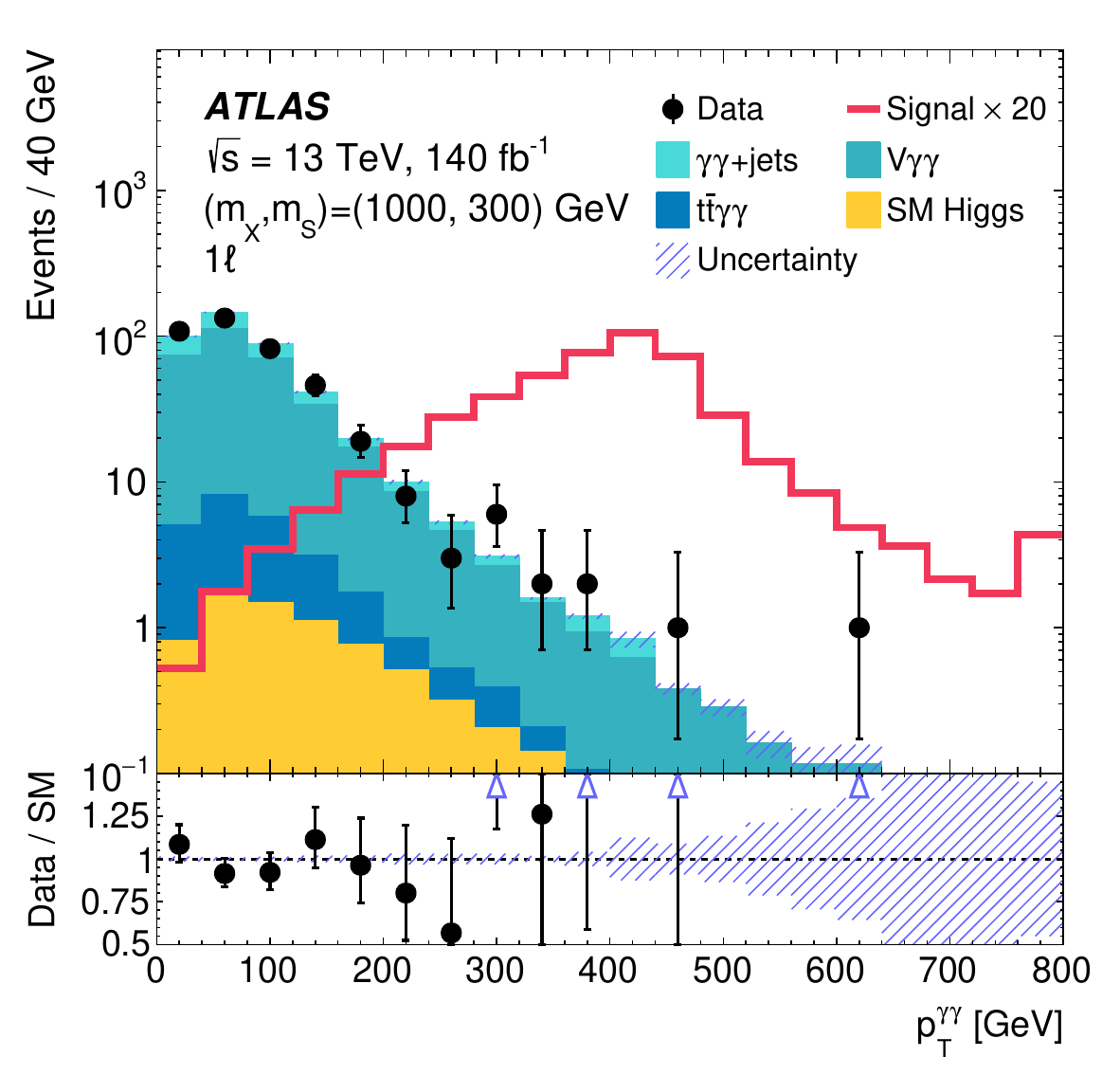}}
\subfigure[]{\includegraphics[width = 0.4\textwidth]{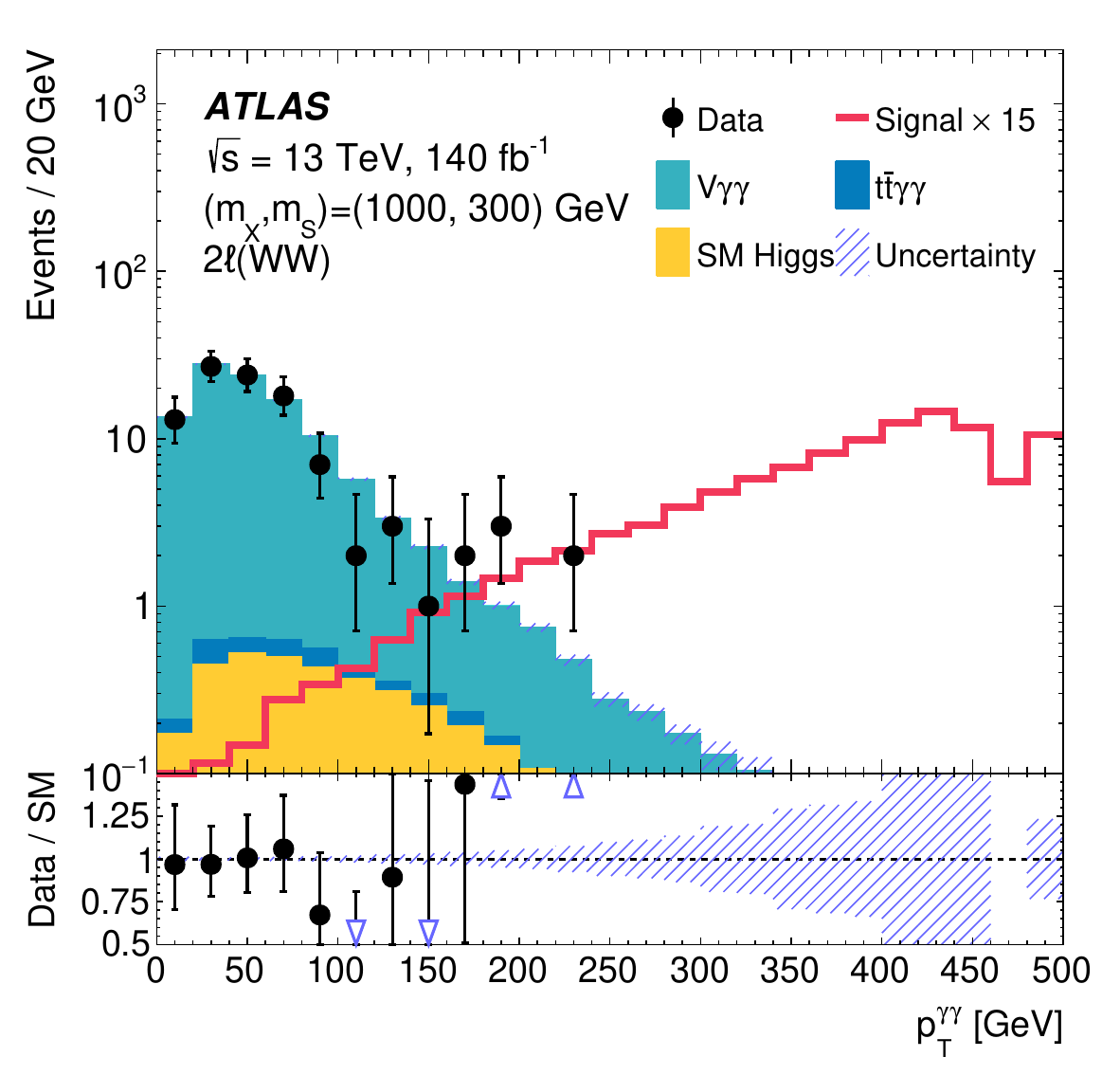}}
\caption{Transverse momentum of the diphoton system, $p_{\textrm{T}}^{\gamma\gamma}$, in the (a) $1\ell$ and (b) $2\ell(WW)$ regions for data and the expected SM background from simulation after the event selection is applied.
The $V\gamma\gamma$ and $\gamma\gamma$+jets simulated background is scaled to match the data yield excluding the $120<m_{\gamma\gamma}<130$~\GeV{} region.
The contribution from the SM single and double Higgs boson processes (denoted ``SM Higgs''), which is estimated from the MC simulation, is also shown.
The $(m_X, \,m_S) = (1000,\, 300)$~\GeV{} signal prediction (open red histogram)  for the scenario of SM-like $\mathscr{B}(S\to WW/ZZ)$ is also shown, normalised to a cross-section corresponding to the 95\% CL upper limit shown in Figure~\ref{fig:limit_scenario_SM}. An additional normalisation factor, as indicated in the legend, is applied to scale the signal for visibility.
The last bin in each distribution contains the overflow.}
\label{fig:pTyy}
\end{center}
\end{figure}

Figure~\ref{fig:BDT} shows the BDT output distributions for data, the expected SM background processes and the $(m_X, \,m_S) = (1000,\, 300)$~\GeV{} signal from simulation.
The BDT output discriminant is used to further split events in loose and tight BDT regions: 1$\ell$loose and 1$\ell$tight regions are defined for the $1\ell$ region, as well as 2$\ell$loose and 2$\ell$tight signal regions for the $2\ell(WW)$ region.
The BDT score threshold values used range from $-0.1$ to $0.2$ depending on the signal mass hypothesis, and result from a scan using the root square of the signal significance in each region added in quadrature: $Z_{\textrm{comb}} = \sqrt{Z_{\textrm{loose}}^2 +  Z_{\textrm{tight}}^2 }$, being

\begin{equation}
Z_{\textrm{loose/tight}} = \sqrt{2 \times \Big[ (s+b) \times \Big( \ln \frac{s+b}{b} \Big) - s \Big]_{\textrm{loose/tight}} }\, ,
\end{equation}

where $s$ represents the signal event yields and $b$ is the background yield in each BDT region.
Both signal and background yields are calculated by considering events in the region of $120<m_{\gamma\gamma}<130$~\GeV{}.
The selected threshold values are established by maximising $Z_{\textrm{comb}}$ under the requirement of the presence of at least one sideband data event in the tight BDT regions.
Table~\ref{tab:cutflow} shows the SM expected event yields, estimated as detailed in Section~\ref{sec:BkgEstimation}, and the observed data for each of the analysis regions.  The expected signal yields for $(m_X,m_S)=(1000,300)$~\GeV{}, considering a $gg\rightarrow X\rightarrow SH$ production cross-section of 1~pb, are provided for comparison.
The $S$ scalar boson is assumed to decay into other SM particles with the same mass-dependent branching ratios of the SM Higgs boson.
\begin{figure}[tbhp]
\begin{center}
\subfigure[]{\includegraphics[width = 0.4\textwidth]{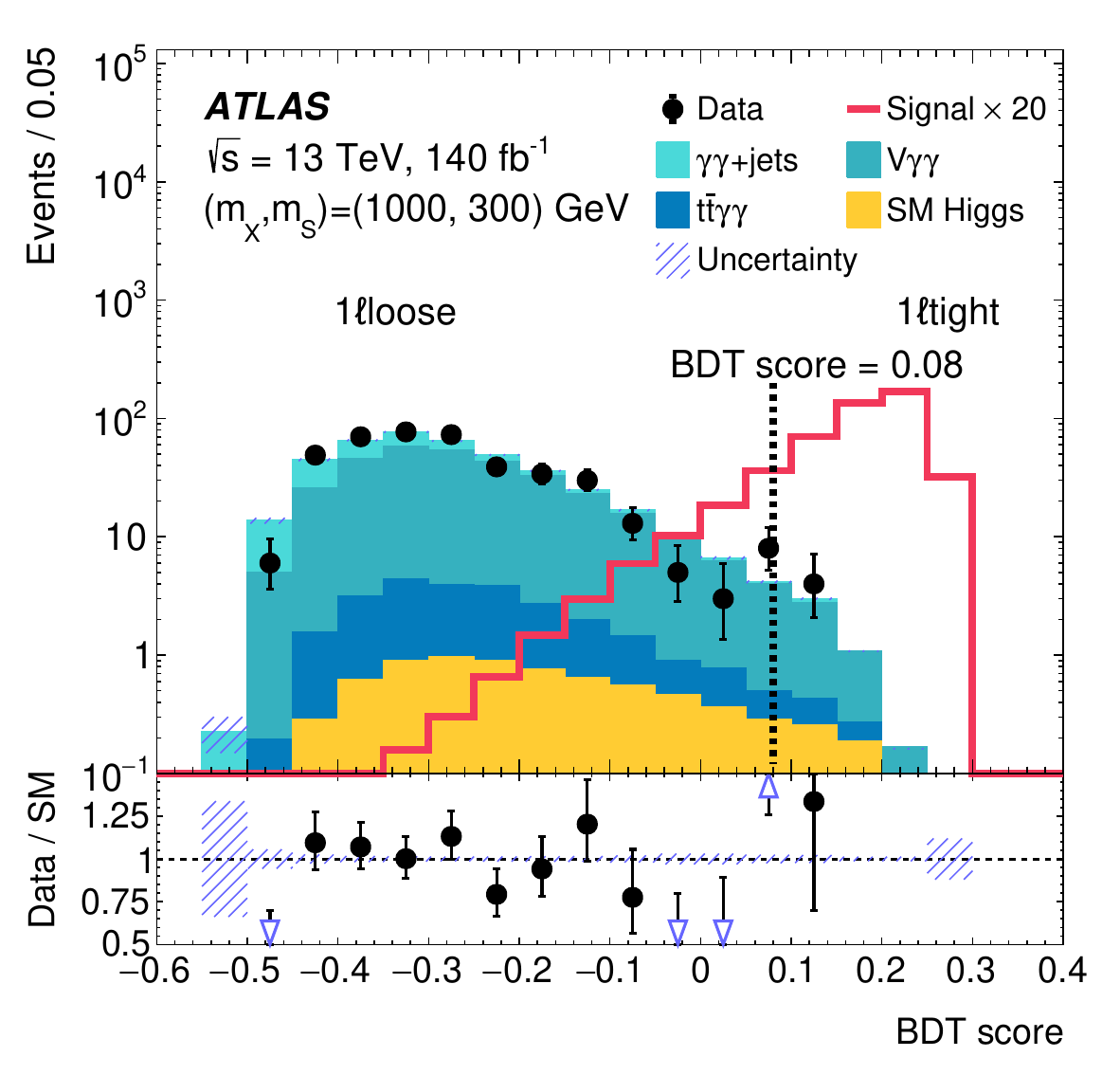}}
\subfigure[]{\includegraphics[width = 0.4\textwidth]{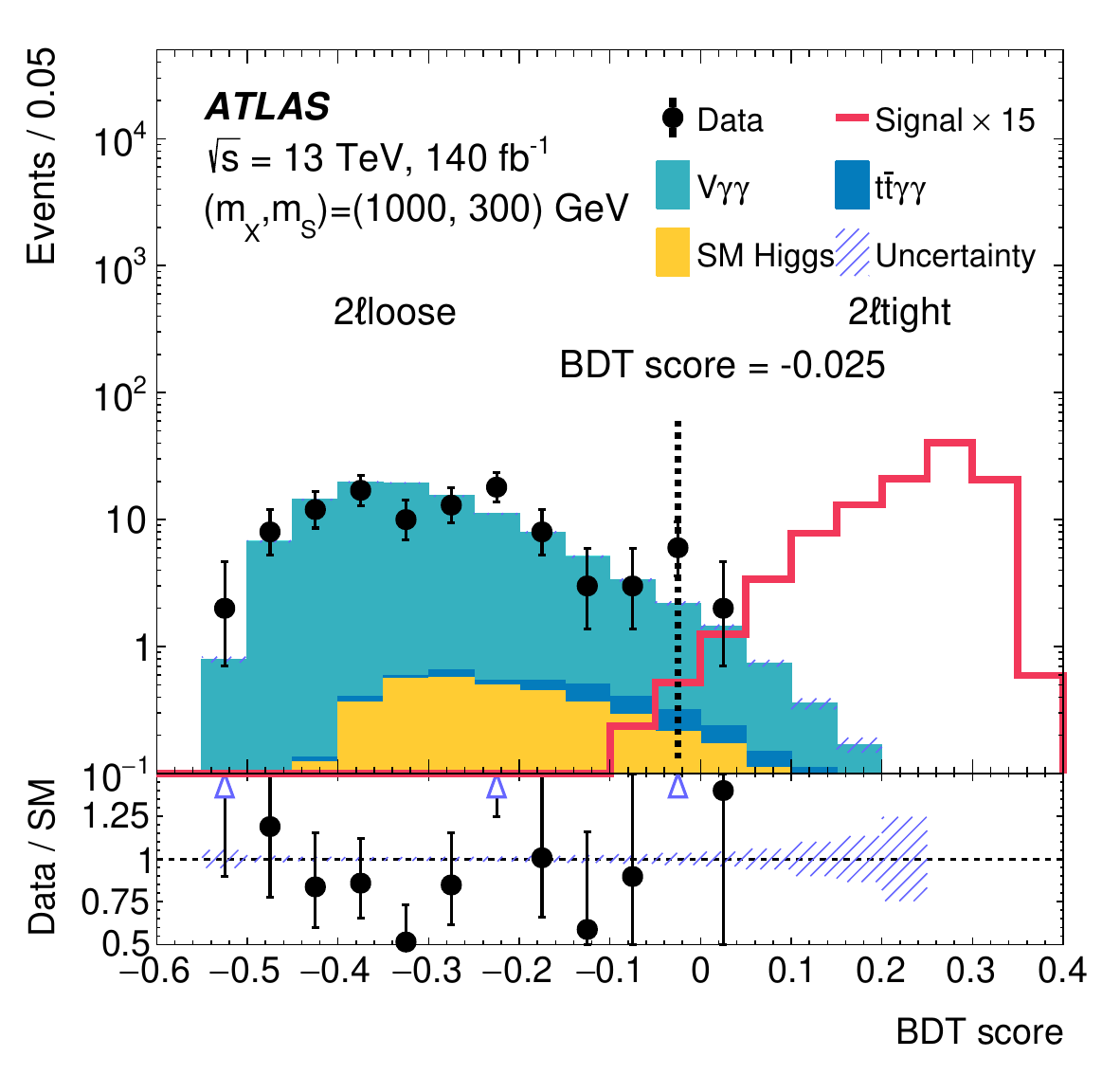}}
\caption{BDT output distributions in the (a) $1\ell$ and (b) $2\ell(WW)$ regions for data and the expected SM background from simulation after the event selection is applied.
The $V\gamma\gamma$ and $\gamma\gamma$+jets simulated background is scaled to match the data yield excluding the $120<m_{\gamma\gamma}<130$~\GeV{} region.
The contribution from the SM single and double Higgs boson processes (denoted ``SM Higgs''), which is estimated from the MC simulation, is also shown.
The $(m_X, \,m_S) = (1000,\, 300)$~\GeV{} signal prediction (open red histogram)  for the scenario of SM-like $\mathscr{B}(S\to WW/ZZ)$ is also shown, normalised to a cross-section corresponding to the 95\% CL upper limit shown in Figure~\ref{fig:limit_scenario_SM}. An additional normalisation factor, as indicated in the legend, is applied to scale the signal for visibility.
The BDT score threshold values are represented by the dashed vertical lines. The shaded band represents the statistical uncertainty on the background prediction.
The last bin in each distribution contains the overflow.}
\label{fig:BDT}
\end{center}
\end{figure}

\begin{table}[t]
\renewcommand{\arraystretch}{1.1}
\begin{center}
\setlength\tabcolsep{0pt}
\caption{Observed data and expected event yields for the different analysis regions after the full selection from Table~\ref{tab:selection} is applied. The continuum background includes the $V\gamma\gamma$, $\gamma\gamma$+jets and $t\bar{t}\gamma\gamma$ processes estimated as described in Section~\ref{sec:BkgEstimation}.
The contribution from the SM single and double Higgs boson processes (denoted  ``SM Higgs'')  is estimated from simulation.
The uncertainties include all sources of systematic uncertainty described in Section~\ref{sec:Systematic}.
Event yields for the $(m_X,m_S)=(1000,300)$~\GeV{} signal are also shown assuming $\sigma(gg\rightarrow X\rightarrow SH)=1$~pb and SM-like $\mathscr{B}(S\to WW/ZZ)$. }
\resizebox{\textwidth}{!}{

\begin{tabular}{l | S[table-format=2.2,table-space-text-pre=x]@{$\;\pm\;$}S[table-format=1.2,table-space-text-post=x]
S[table-format=3.2,table-space-text-pre=x]@{$\;\pm\;$}S[table-format=2.2,table-space-text-post=x] |
S[table-format=2.3,table-space-text-pre=x]@{$\;\pm\;$}S[table-format=1.3,table-space-text-post=x]
S[table-format=3.3,table-space-text-pre=x]@{$\;\pm\;$}S[table-format=2.3,table-space-text-post=x] |
S[table-format=1.3,table-space-text-pre=x]@{$\;\pm\;$}S[table-format=1.3,table-space-text-post=x]
S[table-format=1.3,table-space-text-pre=x]@{$\;\pm\;$}S[table-format=1.3,table-space-text-post=x] }

\hline
\hline
& \multicolumn{8}{c|}{BDT-based regions} & \multicolumn{4}{c}{Cut-based regions} \\
& \multicolumn{2}{c}{1$\ell$tight} & \multicolumn{2}{c}{1$\ell$loose} & \multicolumn{2}{c}{2$\ell$tight} & \multicolumn{2}{c|}{2$\ell$loose} & \multicolumn{2}{c}{$2\ell(ZZ)$} & \multicolumn{2}{c}{$e\mu$}\\

\hline
Continuum  &  6.0  &  2.4  &  405  &  20  &  2.0  &  1.4  &  100  &  10 &  2.0 &  1.4  &  2.0  &  1.4  \\
SM Higgs   &  0.55  &  0.08   &  6.8  &  0.9   &  0.46  &  0.06   &  3.35  &  0.46   &  0.52  &  0.08 &  0.24  &  0.03     \\
\hline
Total background \,  &  6.6  &  2.8  &  412  &  23   &  2.46  &  1.6   &  103  &  11   &  2.52  &  1.6 &  2.24  &  1.5    \\
\hline
\makecell[l]{Signal $(m_X, m_S)$\\ $(1000,300)$~\GeV{}\,  }  &  20.9  &  2.4   &  2.90  &  0.34   &  2.96  &  0.35   &  0.016  &  0.002   &  2.03  &  0.24  & 2.08  &  0.24       \\
\hline
Data   & \multicolumn{2}{c}{6} & \multicolumn{2}{c|}{405} & \multicolumn{2}{c}{2} & \multicolumn{2}{c|}{100} & \multicolumn{2}{c}{2} & \multicolumn{2}{c}{2} \\

\hline
\hline
\end{tabular}
}
\label{tab:cutflow}
\end{center}
\end{table}

%
%
%
%
%
%
%
%
%
%
%
%
%
%
%
%
%
%
%
%
%
%
%
%
%
%
%
%
%
%
%
%
%
%
%
%

%
%
%
%
%
%

%
%
%

%
%
%
%
%
%

%
%
%
%
%
%
%
%
%
%
%
%
%

%
%
%
%

%

%
%
%
%
%
%
%
%
%


%
%

\FloatBarrier
\section{Background estimation}
\label{sec:BkgEstimation}


Background processes can be classified into ``resonant'' or ``continuum'' based on their  $m_{\gamma\gamma}$ spectrum.
The SM Higgs boson single and pair production events form the resonant background component.
These processes are purely estimated from MC simulation.

The continuum background component arises mostly from multi-jet processes associated with two photons ($\gamma\gamma$+jets),
and vector boson or top-antitop-quark production in association with a pair of photons ($V+\gamma\gamma$ and $t\bar{t}\gamma\gamma$).
Contributions from these processes are checked with the MC simulated samples as described in Section~\ref{sec:bkg_sample} and used for the event selection optimisation.
No dedicated MC simulated events are produced for processes with small contribution such as $VV$+$\gamma\gamma$ or processes with jets or leptons misidentified as photons.
Their contributions are included in the data-driven background, which accounts for all possible processes.
The contribution from the continuum background is estimated from a fit to the data $m_{\gamma\gamma}$ distribution in the sideband region with a template.
This template is generated from an analytic function that is obtained from a fit to the $m_{\gamma\gamma}$ distribution in a dedicated data control region due to low statistics of sideband data in the signal region.
These control regions are defined by requiring zero leptons and at least two photons passing looser identification and isolation criteria but failing the signal region photon selections as described in Section~\ref{sec:ObjectSelection}.
Two different control samples are defined based on the number of leptons in each signal region using the selected jet to mimic the lepton behaviour.
The control sample for the $\gamma\gamma + 1\ell$ region selects events with a pair of photons accompanied by one jet.
For the other signal regions with a pair of leptons in the final state, the control sample requires the presence of at least two jets.
A schematic diagram presenting the definitions of different regions is shown in Figure~\ref{fig:region_def}.
The fitted \myy{} shape difference between the signal region and control region is considered as a systematic uncertainty in the background shape estimation.
In addition, the \myy{} shape difference related to the number of leptons in the generator-level is also evaluated and included as a systematic uncertainty.
This uncertainty is estimated by comparing the fit results to the full \myy{} mass range distributions in the control region and the validation region using dedicated lepton-dependence MC samples as described in Section~\ref{sec:bkg_sample},
where the validation region is defined by applying almost the same event selections as the signal region but inverting identification and isolation requirements as same as in the control region.

\begin{figure}[tbhp]
\begin{center}
\includegraphics[width=0.9\textwidth]{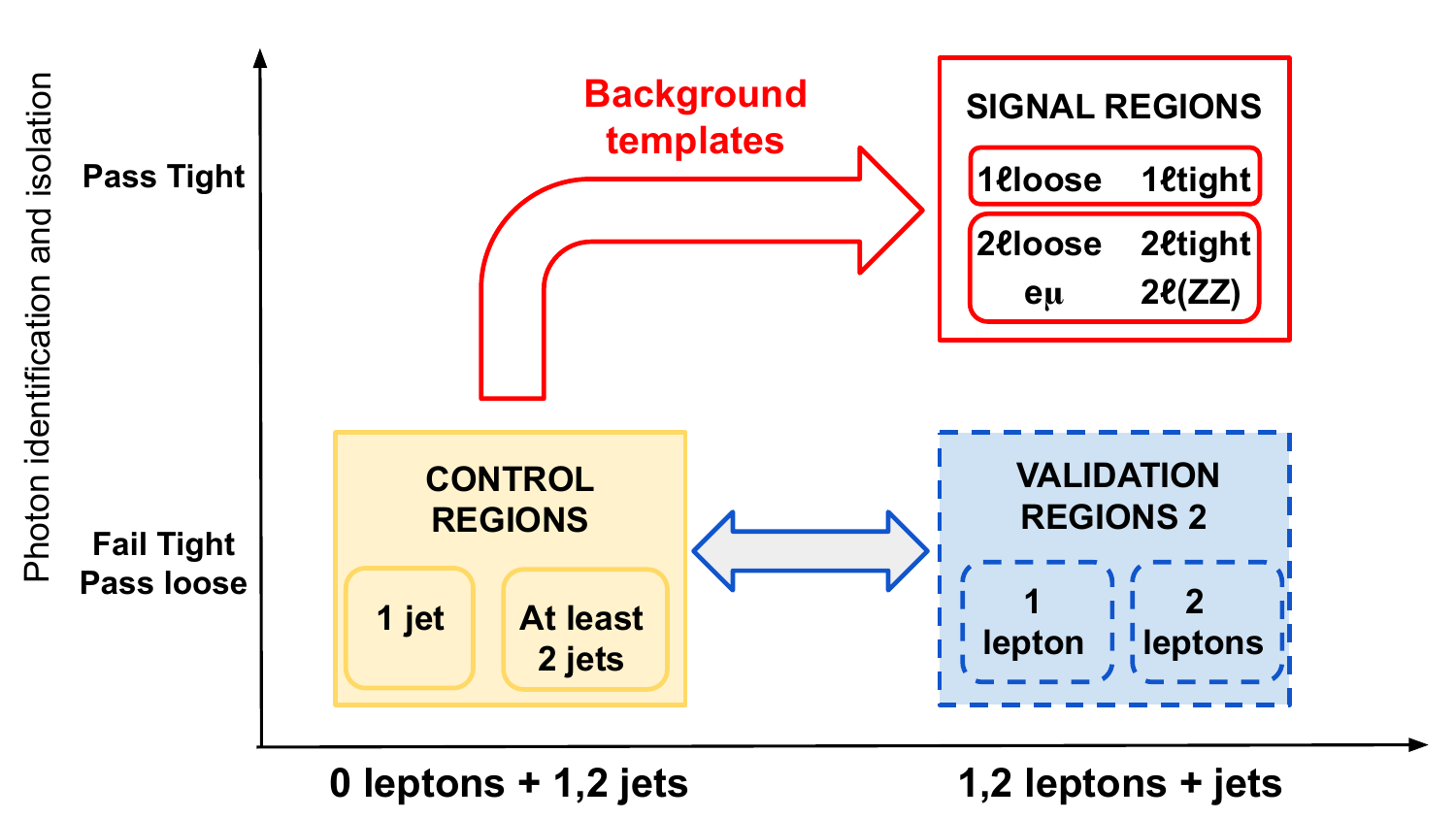}
\end{center}
\caption{Definition of signal and control regions. The red arrow represents the $m_{\gamma\gamma}$ template generated in the 0-lepton control region which is applied to the signal regions. The systematic uncertainty in the background shape, obtained from differences between the control and the validation regions, is indicated by the blue arrow. }
\label{fig:region_def}
\end{figure}

Three types of analytic functions are explored: exponential function, exponential function of a 2$^{\mathrm{nd}}$ order polynomial, and a Chebyshev polynomial of order $n=1,\ldots,5$.
The functional form is chosen via a spurious signal test as described in Ref.~\cite{ATL-PHYS-PUB-2020-028}.
The spurious signal is extracted by performing a signal-plus-background fit to the data \myy{} distribution in the control region (denoted ``background-only'' template),
which is assumed to only include contributions from continuum background processes.
The selection criteria follow the strategy as documented in Ref.~\cite{HIGG-2019-13}.
The spurious signal should be less than 20\% of the background uncertainty. %
In addition to the spurious signal requirement, the goodness of the fit with background functional form to the background-only template is evaluated with a $\chi^{2}$ test and the corresponding $p$-value is required to be greater than 5\%.
When multiple functions pass the criteria, the one with the smallest degrees of freedom is chosen.
The corresponding spurious signal for the selected function is treated as the systematic uncertainty due to background modelling with the analytic function.
In total 80 continuum background functions are estimated corresponding to various signal regions.
Of these 80 functions, 78 are exponential functions of 2$^{\mathrm{nd}}$ order polynomials and the remaining two are simple exponential functions.
Due to the low statistics in the $e\mu$ and $2\ell(ZZ)$ signal regions and the fact that the $\myy$ distribution shows a negligible dependence on the flavour of leptons,
the continuum background shape estimated in the 2$\ell$tight region is also applied to these two regions.

%
%


\FloatBarrier
\section{Systematic uncertainties}
\label{sec:Systematic}

%
Systematic uncertainties arise from the theory modelling of signal and background, the detector simulation and instrumental effects, and the estimation of the continuum background.

\subsection{Theoretical uncertainties}
Theoretical uncertainties are considered for signal and the SM single and double Higgs boson backgrounds.
Uncertainties from six sources are considered: from PDF set and strong coupling constant $\alpha_{S}$, from the QCD factorisation and renormalisation scales ($\mu_{\mathrm{F}}$ and $\mu_{\mathrm{R}}$), and from the parton shower parameters and hadronisation models.

To evaluate the impact of varying the PDF set choice and $\alpha_{S}$ value, event weights corresponding to alternative PDF sets and $\alpha_{S}$ values are generated for each event along with the nominal weight.
Effects on signal region yields are considered as systematic uncertainties. Variations on signal and SM Higgs boson yields are found to be 6\% and 4\% respectively.

The systematic uncertainty due to higher-order QCD effects is estimated by independently varying the QCD factorisation and renormalisation scales up and down from their nominal values by a factor of two, taking the envelope of the 7-point variation.
Their impacts on signal and SM Higgs boson background yields are about 9\% and 13\% respectively.

The uncertainty due to the parton shower and hadronisation model is estimated by comparing the yields from the nominal MC samples using \PYTHIA[8], with alternative samples using instead \HERWIG[7].
The corresponding uncertainties in the signal and Higgs boson background yields are 5\% and 3\% respectively.

\subsection{Experimental uncertainties}

Systematic uncertainties arising from the luminosity determination, pileup modelling, and trigger, reconstruction and selection efficiencies, as well as energy scales and resolutions, are considered for signal, single and double Higgs boson events.
Their impacts on both the normalisation and shape are included in the statistical analysis for the final results.

The uncertainty in the integrated luminosity for 2015--2018 data taking period is 0.83\%~\cite{DAPR-2021-01},
obtained using the LUCID-2 detector for the primary luminosity measurement, complemented by measurements using the inner detector and calorimeters.

The uncertainty in the modelling of the pileup distribution in the simulation is estimated to have 2\%--3\% impacts on yields of signal, single and double Higgs boson events.
The photon reconstruction, identification and isolation efficiencies are measured using three data-driven techniques as mentioned in Ref.~\cite{EGAM-2021-01}. %
Their effects on yields and shapes are estimated by varying the measured efficiency scale factors between data and simulation, resulting in less than 2\% variations for signal yields.
Uncertainties in the photon energy scale and resolution described in Ref.~\cite{EGAM-2018-01} are considered as well.
These uncertainties affect the signal yield less than 0.5\%; a similar impact is found for single Higgs boson and double Higgs bosons events.
The uncertainty in the photon trigger efficiency is also considered and its impact on event yields is found to be negligible.

Uncertainties in the electron reconstruction, identification and isolation efficiencies are reported in Ref.~\cite{EGAM-2021-01}.
They affect the signal and SM Higgs boson yields by about 2\%.
Uncertainties in electron energy scale and resolution are also evaluated and found to have a negligible impact.

In addition, uncertainties in muon reconstruction, identification, isolation efficiencies as well as the muon momentum scale and resolution~\cite{MUON-2018-03},
and uncertainties in the jet energy scale and resolution~\cite{JETM-2018-05} are also considered.
Furthermore, uncertainties arising from jet property selections: jet-vertex-tagger~\cite{PERF-2014-03} and $b$-jet tagging~\cite{FTAG-2018-01, FTAG-2020-08, FTAG-2019-02} are included.
Finally, the uncertainty related to \MET{} resulting from tracks not associated with the selected objects~\cite{JETM-2020-03} is considered. %
All these uncertainties were found to have a negligible impact on the signal and SM Higgs boson yields.

\subsection{Continuum background modelling uncertainty}
The continuum background estimation (see Section~\ref{sec:BkgEstimation}) assumes no significant shape differences between the control region (events with no leptons) and the signal region (events with at least one lepton).
An uncertainty (called lepton-dependence uncertainty) associated with the background modelling is evaluated by comparing the shape of the $\myy$ distribution in $\gamma\gamma+0\ell$ events with that in $\gamma\gamma+\ell \nu$jj and $\gamma\gamma+\ell \nu \ell \nu$ events from dedicated MC simulated samples (see Section~\ref{sec:bkg_sample}).
The variations between diphoton events with and without leptons are computed using the $m_{\gamma\gamma}$ distribution in the $[105, 160]$~\GeV{} range.
The average variation over all bins is about 2\% for all signal regions.
In addition to lepton-dependence uncertainty, a ~2\% uncertainty arises from the comparison of the shape of the $\myy$ distribution in the control and the signal regions.
The systematic uncertainty arising from a potential bias from the background shape functional form choice is accounted by the spurious signal uncertainty (see Section~\ref{sec:BkgEstimation}).


\FloatBarrier
\section{Results}
\label{sec:Result}

The contribution of a potential signal in the data is extracted through a simultaneous fit to the \myy{} distributions in all six signal regions,
which is implemented with the \textsc{RooFit}~\cite{RooFit} and \textsc{RooStats}~\cite{roostats} frameworks.
The fit is performed with a binned likelihood model built from the product of the Poission distribution in each bin and region,
and including Gaussian distributions to describe the effect of systematic uncertainties.
For each \myy{} distribution, 22 equal width bins in a range of 105~\GeV{} to 160~\GeV{} are used in the fit.
The parameter-of-interest is $\sigma(gg\to X)\times \mathscr{B}(X\to SH)$ and is left unconstrained in the fit.
The shape of the signal for each individual region is obtained from simulated events.
Contributions from the single and double Higgs boson processes are estimated from MC simulated samples with their normalisation fixed to SM predictions.
Theoretical and experimental uncertainties corresponding to both the signal and the Higgs boson backgrounds are included in the fit and controlled by the nuisance parameters.
For the continuum background, the shape is estimated with the method described in Section~\ref{sec:BkgEstimation} and kept fixed in the fit, while its normalisation is left unconstrained.
In total four individual normalisation factors are included corresponding to 1$\ell$tight, 1$\ell$loose, 2$\ell$tight and 2$\ell$loose, respectively.
The $e\mu$ and $2\ell(ZZ)$ regions share the same normalisation factor for continuum background as the 2$\ell$tight region.
Background shape uncertainty and the spurious signal uncertainty are also considered in the fit and described by the corresponding nuisance parameters.
To improve the robustness of the fit, any systematic uncertainty with less than 0.5\% impact on either shape or yield is removed from the fit model.
With such requirement, only uncertainties related to photon and electron, pileup reweighting, theory, and continuum background estimation are left, and others are ignored during the fit.

Three scenarios are considered for the final results corresponding to three hypotheses
on the $S\to WW/ZZ$ branching ratios: SM-like  $\mathscr{B}(S\to WW/ZZ)$~\cite{deFlorian:2016spz}, $\mathscr{B}(S\to WW)=100\%$, and $\mathscr{B}(S\to ZZ)=100\%$.

Figure~\ref{fig:postfit_X1000_S300} presents the \myy{} distributions in the six signal regions after performing the signal-plus-background fit corresponding to a signal with $m_{X}=1000\,\GeV$ and %
$m_{S}=300\,\GeV$ for the SM-like  $\mathscr{B}(S\to WW/ZZ)$ scenario.
Other branching-ratio scenarios and signal-mass hypotheses are also tested and no deviation with respect to the SM background expectation is observed.
Consequently, 95\% CL upper limits are set on $\sigma(gg\to X)\times \mathscr{B}(X\to SH)$ for each branching-ratio scenario and signal-mass hypothesis using the profile likelihood ratio technique with the asymptotic approximation~\cite{Asymptotic} and the $CL_s$~\cite{CLs1,CLs2} method.
The results are validated using pseudo-experiments and found to agree within 20\%.

Figure~\ref{fig:limit_scenario_SM} shows the 95\% CL observed and expected limits on $\sigma(gg\to X)\times \mathscr{B}(X\to SH)$ as a function of $m_{X}$ and $m_{S}$ for the SM-like  $\mathscr{B}(S\to WW/ZZ)$ scenario.
The observed (expected) limit ranges from 530~fb (800~fb) to 120~fb (170~fb) depending on the scalar masses.
The results are dominated by the $1\ell$tight region, with the other regions contributing with comparably lower sensitivity.
In the $1\ell$tight region, a slight deficit in the data yield compared to the background expectation is observed across all mass hypotheses,
which leads to a better observed limit than the expected one for every mass point.

\begin{figure}[tbh]
\begin{center}
\subfigure[]{\includegraphics[width=0.33\textwidth]{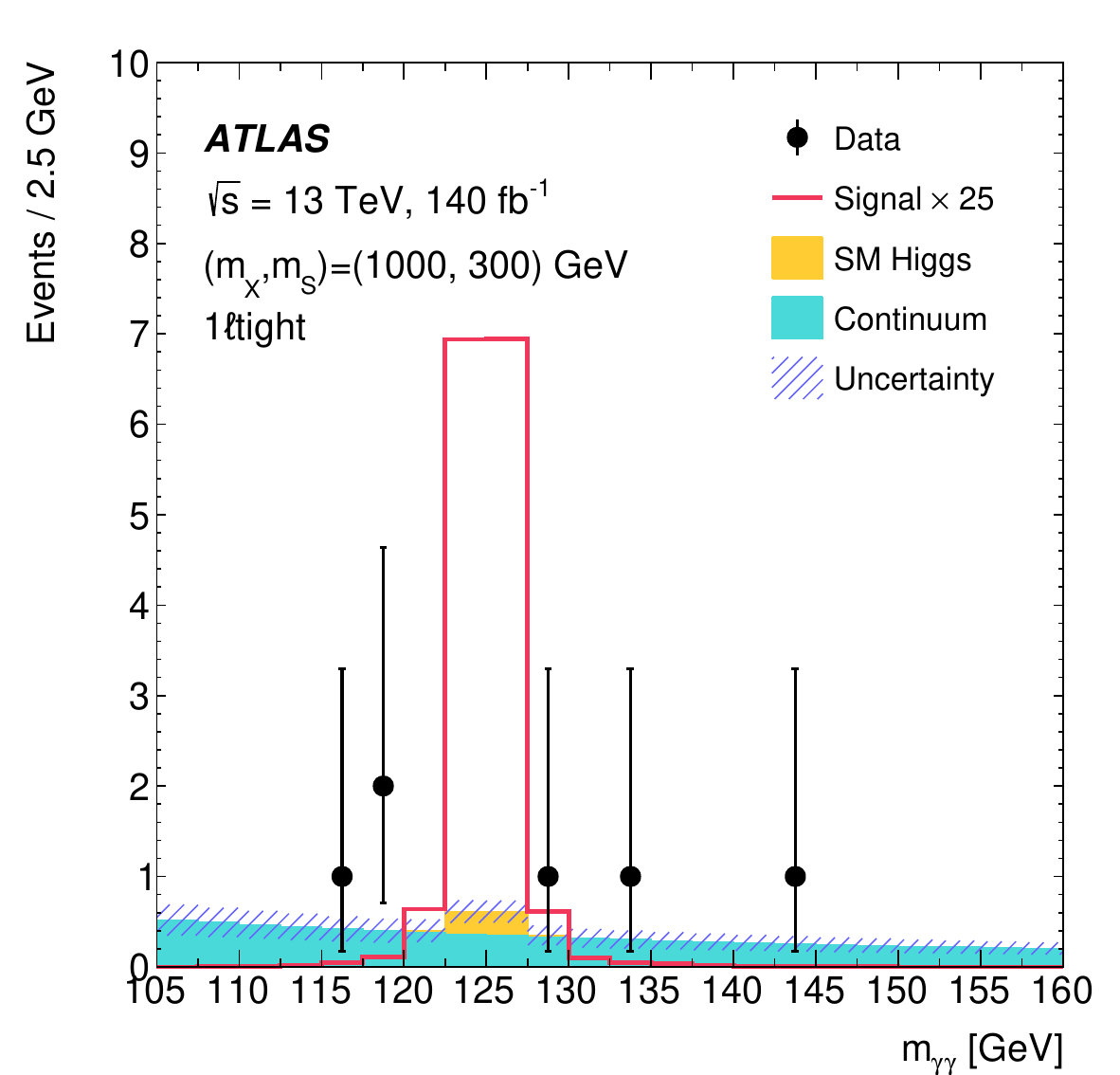}}
\subfigure[]{\includegraphics[width=0.33\textwidth]{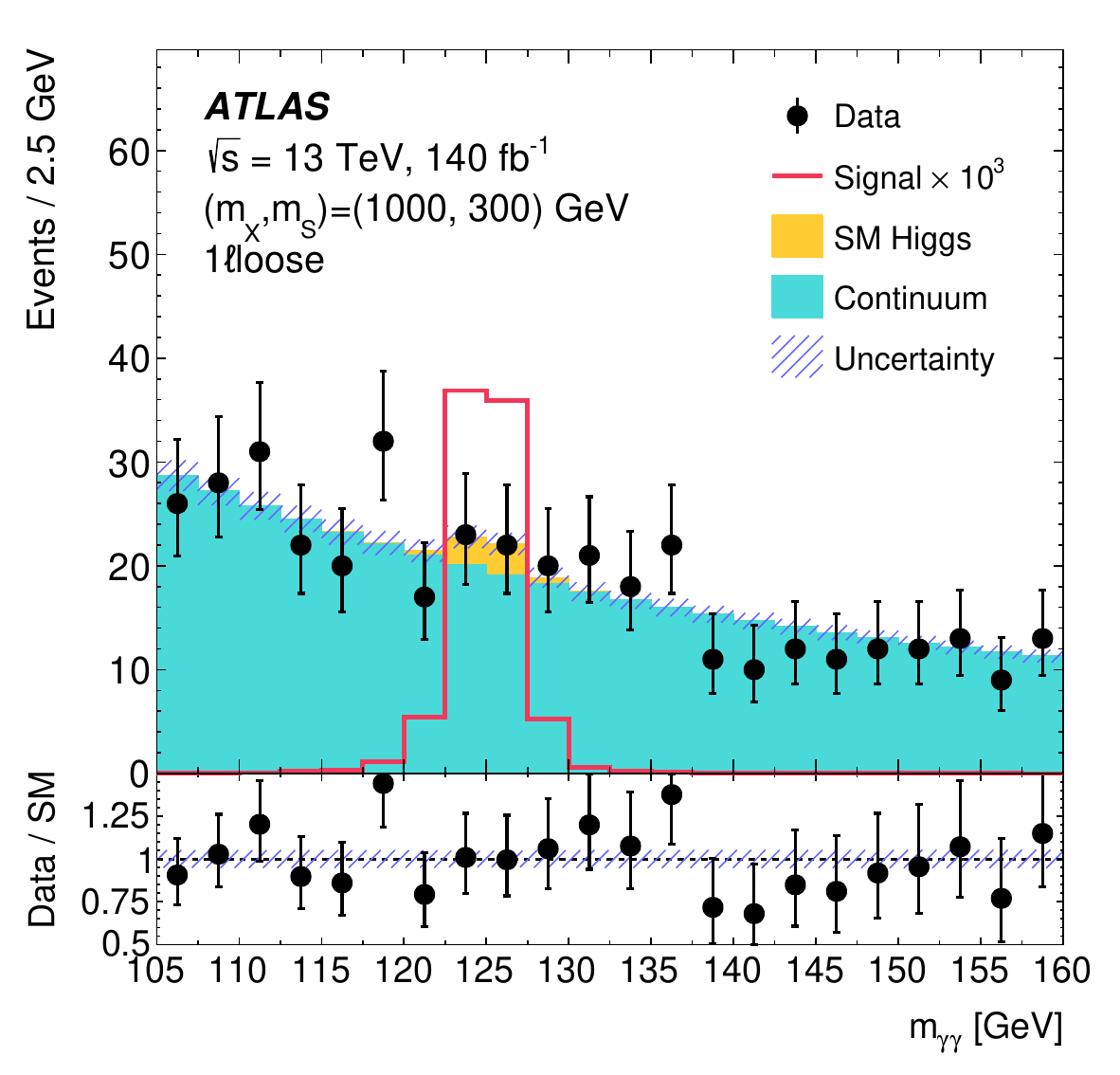}}
\subfigure[]{\includegraphics[width=0.33\textwidth]{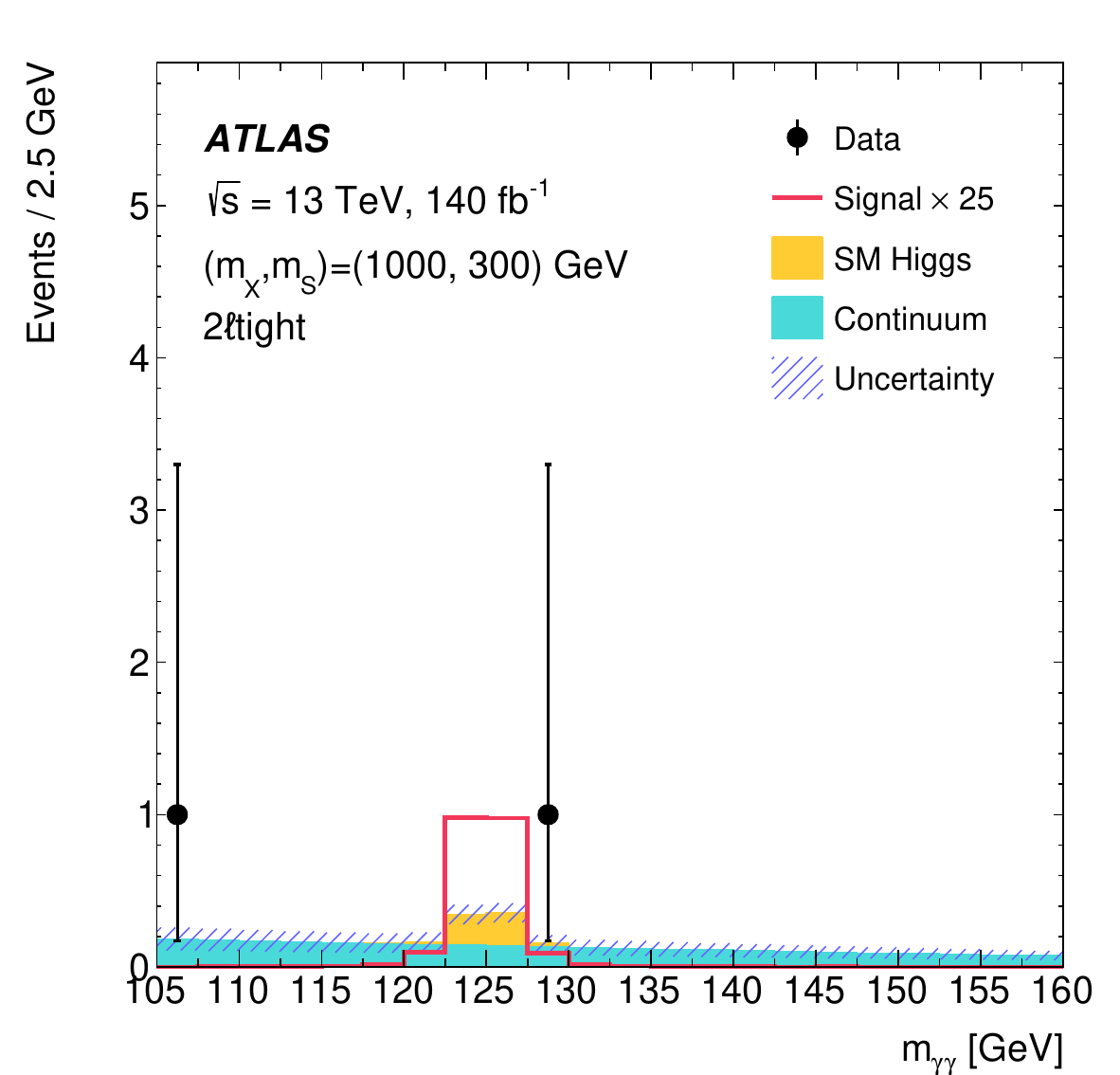}}
\subfigure[]{\includegraphics[width=0.33\textwidth]{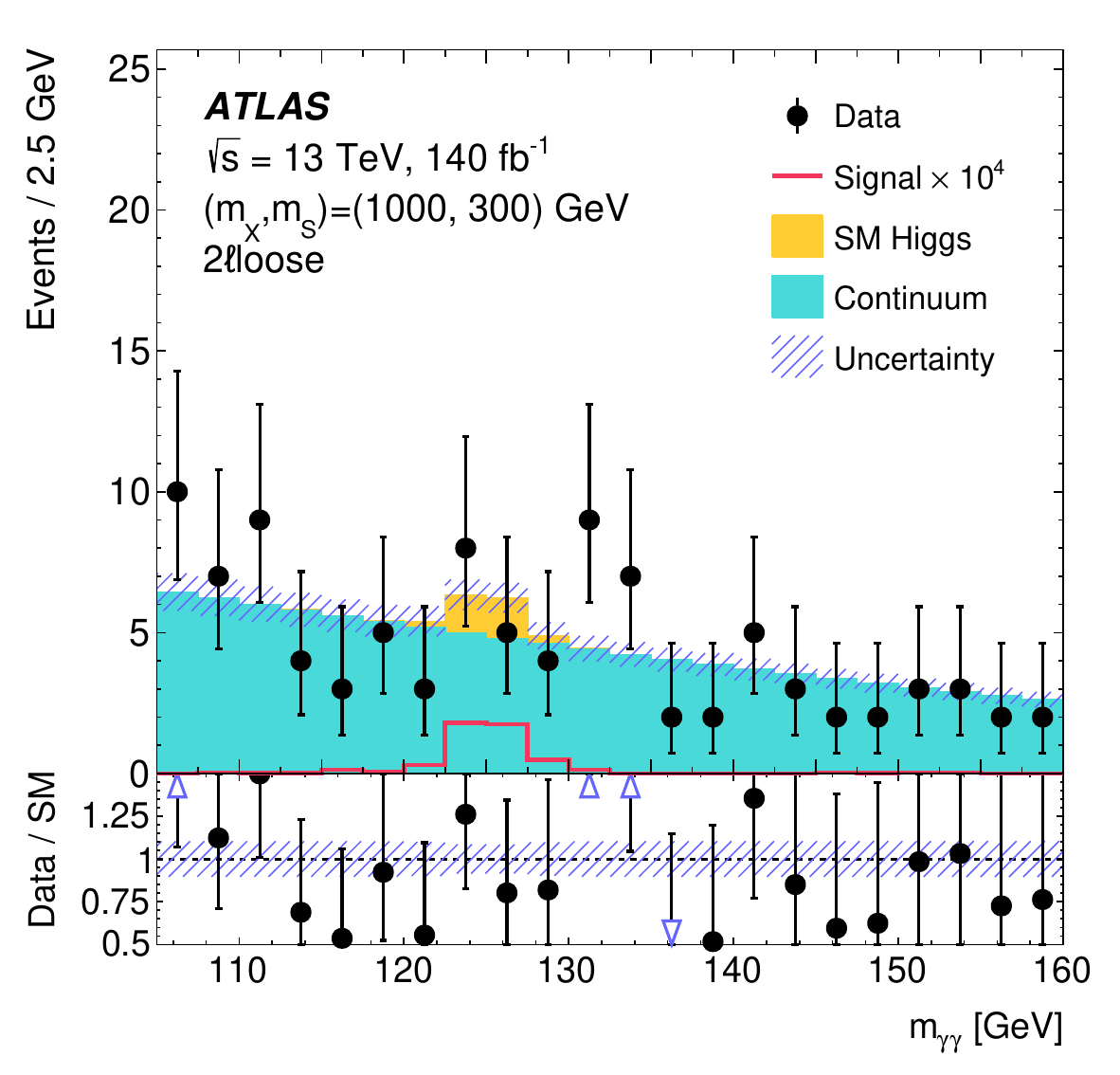}}
\subfigure[]{\includegraphics[width=0.33\textwidth]{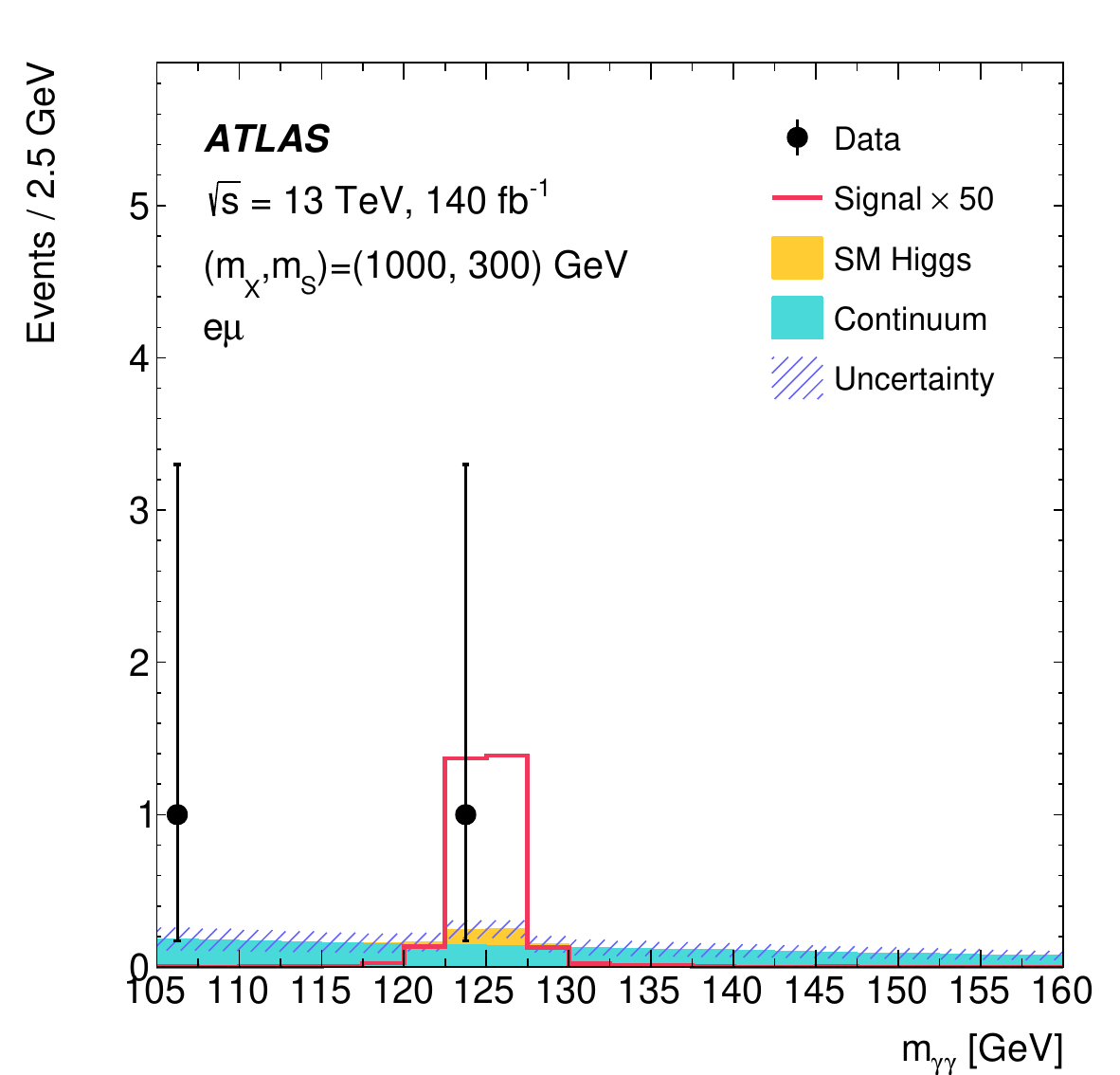}}
\subfigure[]{\includegraphics[width=0.33\textwidth]{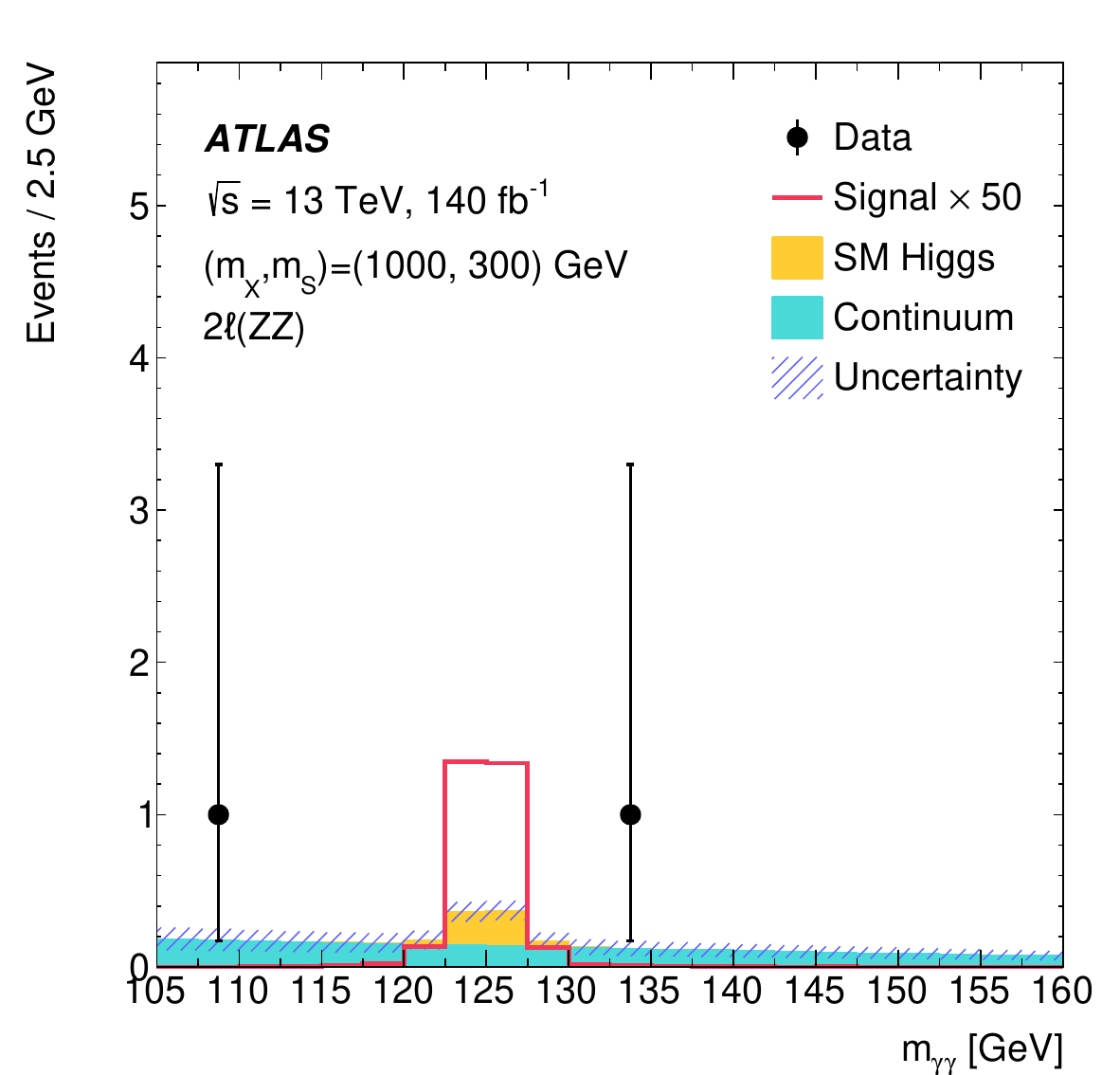}}
\caption{Distribution of  \myy{} after the signal-plus-background fit to data in the (a) 1$\ell$tight, (b) 1$\ell$loose, (c) 2$\ell$tight, (d) 2$\ell$loose, (e) $e\mu$ and (f) $2\ell(ZZ)$ regions.
The contribution from the SM single and double Higgs boson processes (denoted ``SM Higgs''), which is estimated from the MC simulation, is shown added on top of the continuum background distribution.
The $(m_X, \,m_S) = (1000,\, 300)$~\GeV{} signal prediction (open red histogram)  for the scenario of SM-like $\mathscr{B}(S\to WW/ZZ)$ is also shown, normalised to a cross-section corresponding to the 95\% CL upper limit shown in Figure~\ref{fig:limit_scenario_SM}.
An additional normalisation factor, as indicated in the legend, is applied to scale the signal for visibility.
}
\label{fig:postfit_X1000_S300}
\end{center}
\end{figure}

\clearpage

\begin{figure}[tb]
\begin{center}
\includegraphics[width=0.80\textwidth]{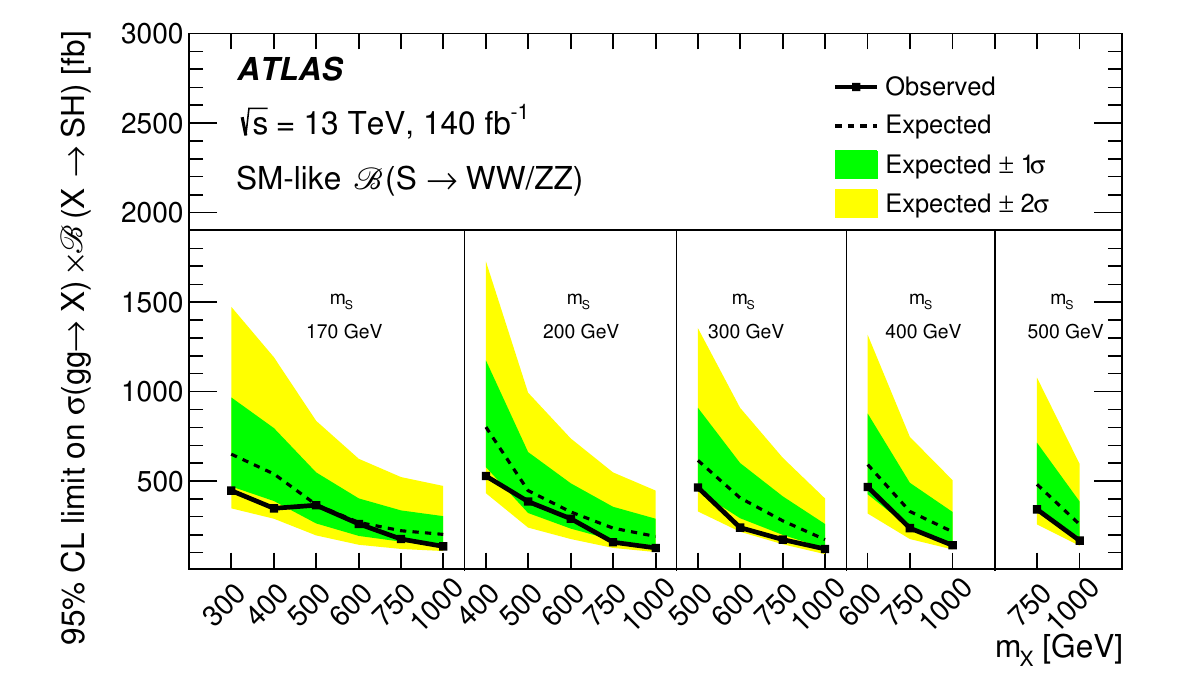}
\caption{Observed (solid line) and expected (dashed line) 95\% CL upper limits on $\sigma(gg\to X)\times \mathscr{B}(X\to SH)$  as a function of $m_X$ and $m_S$, under the assumption of SM-like $\mathscr{B}(S\to WW/ZZ)$. The green and yellow shaded areas indicate the $\pm 1$  and $\pm 2$ standard deviations around the expected limit. }
\label{fig:limit_scenario_SM}
\end{center}
\end{figure}

Limits corresponding to assumptions of the scalar $S$ with a 100\% decay branching ratio to $WW$ or $ZZ$ are derived and presented in Figures~\ref{fig:limit_scenario_HWW} and \ref{fig:limit_scenario_HZZ}.
Under the assumption that $\mathscr{B}(S\to WW)=100\%$, the observed limit varies from 470~fb to 91~fb whereas the expected limit ranges from 610~fb to 120~fb.
The upper limits under the scenario $\mathscr{B}(S\to ZZ)=100\%$ are significantly higher: from 1530~fb to 360~fb for observed limits and from 2160~fb to 510~fb for expected limits.
The analysis sensitivity is limited by the statistical uncertainty, with systematic uncertainties degrading the expected limits by about 2\%.

These results are comparable to the $X\rightarrow SH \rightarrow VV\tau\tau$~\cite{HDBS-2022-44} search, which observes an upper limit on the production cross-section from 540~fb to 72~fb, assuming SM-like  $\mathscr{B}(S\to WW/ZZ)$.
Under the $\mathscr{B}(S\to WW)=100\%$ and $\mathscr{B}(S\to ZZ)=100\%$ scenarios, the upper limits on the production cross-section and decay branching ratio are in the ranges 26 -- 3~fb and 33 -- 6~fb, respectively.
By correcting by the $\mathscr{B}(H\rightarrow \tau\tau)$, these results can be expressed in upper limits on the $X\rightarrow SH$ production cross-section, and compared to those obtained by this analysis.
These upper limits are set in the ranges 410 -- 47~fb and 520 -- 95~fb for $\mathscr{B}(S\rightarrow WW)=100\%$ and $\mathscr{B}(S\rightarrow ZZ)=100\%$, respectively.

\begin{figure}[tb]
\begin{center}
\includegraphics[width=0.80\textwidth]{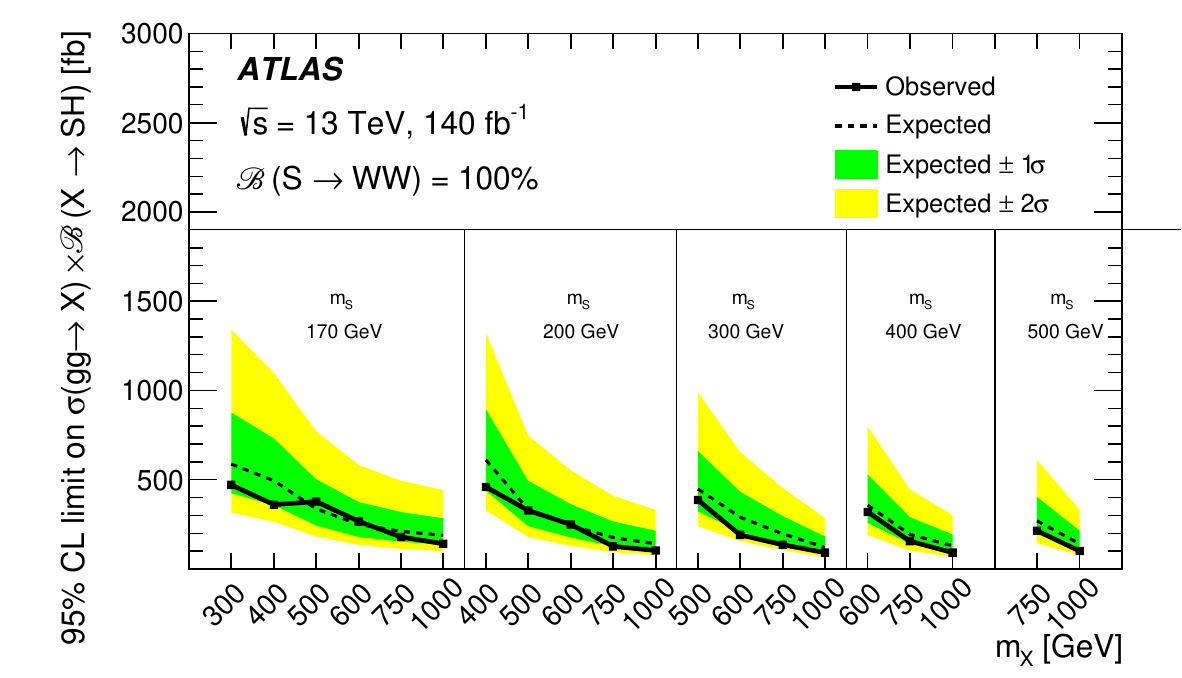}
\caption{Observed (solid line) and expected (dashed line) 95\% CL upper limits on $\sigma(gg\to X)\times \mathscr{B}(X\to SH)$  as a function of $m_X$ and $m_S$, under the assumption of $\mathscr{B}(S\to WW)=100\%$. The green and yellow shaded areas indicate the $\pm 1$  and $\pm 2$ standard deviations around the expected limit. }
\label{fig:limit_scenario_HWW}
\end{center}
\end{figure}

\begin{figure}[tb]
\begin{center}
\includegraphics[width=0.80\textwidth]{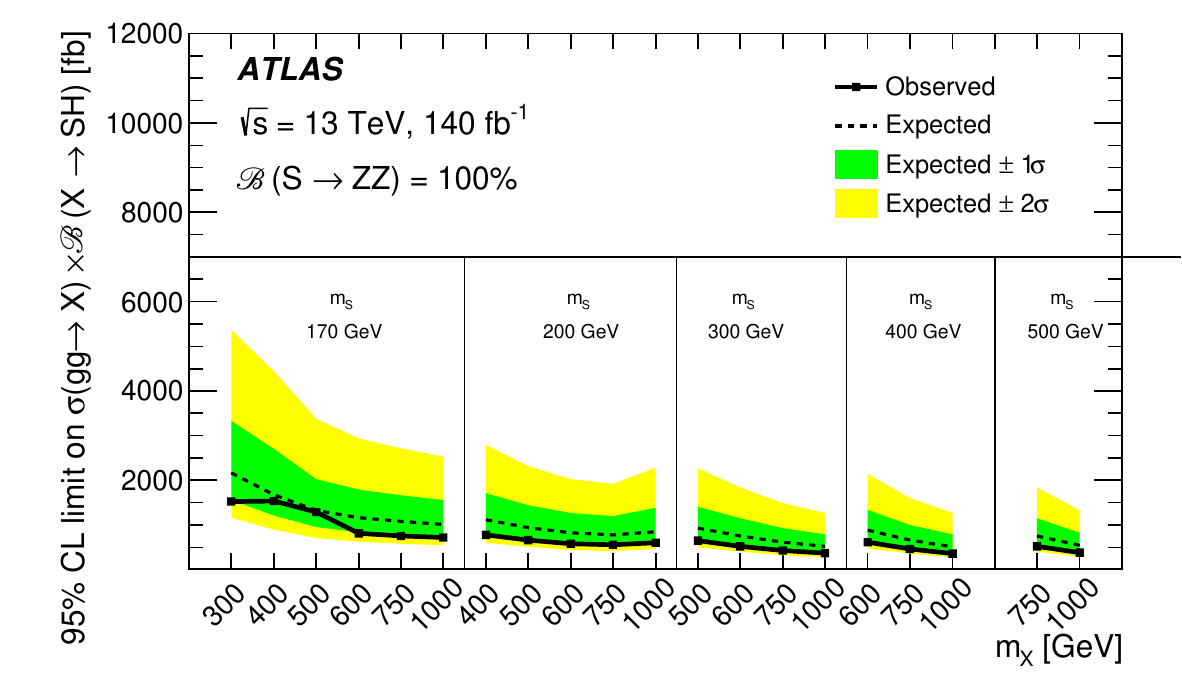}
\caption{Observed (solid line) and expected (dashed line) 95\% CL upper limits on $\sigma(gg\to X)\times \mathscr{B}(X\to SH)$  as a function of $m_X$ and $m_S$, under the assumption of $\mathscr{B}(S\to ZZ)=100\%$. The green and yellow shaded areas indicate the $\pm 1$  and $\pm 2$ standard deviations around the expected limit. }
\label{fig:limit_scenario_HZZ}
\end{center}
\end{figure}


%

%
%
\FloatBarrier

\section{Conclusion}
\label{sec:Conclusion}

%

This paper presents the first search for the $X\rightarrow SH \rightarrow VV\gamma\gamma$ process by selecting events with a pair of photons accompanied by one or two leptons (electrons or muons).
The analysis is based on 140~fb$^{-1}$ of proton--proton collision data at $\sqrt{s}=13$~\TeV{} recorded with the ATLAS detector at the LHC.
The $X\to SH\to VV\gamma\gamma$ signal is searched for over the $300 \leq m_X \leq 1000$~GeV and $170 \leq m_S \leq 500$~GeV mass ranges, probing lower $m_X$ values than the ATLAS $SH\to VV\tau\tau$ search, and complementing the ATLAS $X\to SH \to bb\gamma\gamma$ search by testing a different $S$-boson decay mode.

No excess of events above the expected SM background is observed and 95\% CL upper limits are set on the cross-section times branching ratio, $\sigma(gg\to X)\times \mathscr{B}(X\to SH)$, under different assumptions for the $S\to WW/ZZ$ branching ratios.
The observed (expected) upper limits lie in the range of 530\,--\,120~fb (800\,--\,170~fb) under the assumption that $\mathscr{B}(S\to WW/ZZ)$ corresponding to those the SM Higgs boson would have at the mass of the $S$ particle.
The corresponding observed (expected) upper limits on the cross-section are in the range of 470\,--\,91~fb (610\,--\,120~fb) under the assumption of $\mathscr{B}(S\rightarrow WW)=100\%$.
Alternatively, under the assumption of $\mathscr{B}(S\to ZZ)=100\%$, the observed (expected) limits are in the range of 1530\,--\,360~fb (2160\,--\,510~fb).

%
%
%
%

%


%

%

%
\section*{Acknowledgements}
%

%
%

%
%

We thank CERN for the very successful operation of the LHC and its injectors, as well as the support staff at
CERN and at our institutions worldwide without whom ATLAS could not be operated efficiently.

The crucial computing support from all WLCG partners is acknowledged gratefully, in particular from CERN, the ATLAS Tier-1 facilities at TRIUMF/SFU (Canada), NDGF (Denmark, Norway, Sweden), CC-IN2P3 (France), KIT/GridKA (Germany), INFN-CNAF (Italy), NL-T1 (Netherlands), PIC (Spain), RAL (UK) and BNL (USA), the Tier-2 facilities worldwide and large non-WLCG resource providers. Major contributors of computing resources are listed in Ref.~\cite{ATL-SOFT-PUB-2023-001}.

We gratefully acknowledge the support of ANPCyT, Argentina; YerPhI, Armenia; ARC, Australia; BMWFW and FWF, Austria; ANAS, Azerbaijan; CNPq and FAPESP, Brazil; NSERC, NRC and CFI, Canada; CERN; ANID, Chile; CAS, MOST and NSFC, China; Minciencias, Colombia; MEYS CR, Czech Republic; DNRF and DNSRC, Denmark; IN2P3-CNRS and CEA-DRF/IRFU, France; SRNSFG, Georgia; BMBF, HGF and MPG, Germany; GSRI, Greece; RGC and Hong Kong SAR, China; ISF and Benoziyo Center, Israel; INFN, Italy; MEXT and JSPS, Japan; CNRST, Morocco; NWO, Netherlands; RCN, Norway; MNiSW, Poland; FCT, Portugal; MNE/IFA, Romania; MESTD, Serbia; MSSR, Slovakia; ARIS and MVZI, Slovenia; DSI/NRF, South Africa; MICIU/AEI, Spain; SRC and Wallenberg Foundation, Sweden; SERI, SNSF and Cantons of Bern and Geneva, Switzerland; NSTC, Taipei; TENMAK, T\"urkiye; STFC/UKRI, United Kingdom; DOE and NSF, United States of America.

Individual groups and members have received support from BCKDF, CANARIE, CRC and DRAC, Canada; CERN-CZ, FORTE and PRIMUS, Czech Republic; COST, ERC, ERDF, Horizon 2020, ICSC-NextGenerationEU and Marie Sk{\l}odowska-Curie Actions, European Union; Investissements d'Avenir Labex, Investissements d'Avenir Idex and ANR, France; DFG and AvH Foundation, Germany; Herakleitos, Thales and Aristeia programmes co-financed by EU-ESF and the Greek NSRF, Greece; BSF-NSF and MINERVA, Israel; NCN and NAWA, Poland; La Caixa Banking Foundation, CERCA Programme Generalitat de Catalunya and PROMETEO and GenT Programmes Generalitat Valenciana, Spain; G\"{o}ran Gustafssons Stiftelse, Sweden; The Royal Society and Leverhulme Trust, United Kingdom.

In addition, individual members wish to acknowledge support from Armenia: Yerevan Physics Institute (FAPERJ); CERN: European Organization for Nuclear Research (CERN PJAS); Chile: Agencia Nacional de Investigaci\'on y Desarrollo (FONDECYT 1230812, FONDECYT 1230987, FONDECYT 1240864); China: Chinese Ministry of Science and Technology (MOST-2023YFA1605700), National Natural Science Foundation of China (NSFC - 12175119, NSFC 12275265, NSFC-12075060); Czech Republic: Czech Science Foundation (GACR - 24-11373S), Ministry of Education Youth and Sports (FORTE CZ.02.01.01/00/22\_008/0004632), PRIMUS Research Programme (PRIMUS/21/SCI/017); EU: H2020 European Research Council (ERC - 101002463); European Union: European Research Council (ERC - 948254, ERC 101089007), Horizon 2020 Framework Programme (MUCCA - CHIST-ERA-19-XAI-00), European Union, Future Artificial Intelligence Research (FAIR-NextGenerationEU PE00000013), Italian Center for High Performance Computing, Big Data and Quantum Computing (ICSC, NextGenerationEU); France: Agence Nationale de la Recherche (ANR-20-CE31-0013, ANR-21-CE31-0013, ANR-21-CE31-0022, ANR-22-EDIR-0002), Investissements d'Avenir Labex (ANR-11-LABX-0012); Germany: Baden-Württemberg Stiftung (BW Stiftung-Postdoc Eliteprogramme), Deutsche Forschungsgemeinschaft (DFG - 469666862, DFG - CR 312/5-2); Italy: Istituto Nazionale di Fisica Nucleare (ICSC, NextGenerationEU), Ministero dell'Università e della Ricerca (PRIN - 20223N7F8K - PNRR M4.C2.1.1); Japan: Japan Society for the Promotion of Science (JSPS KAKENHI JP22H01227, JSPS KAKENHI JP22H04944, JSPS KAKENHI JP22KK0227, JSPS KAKENHI JP23KK0245); Netherlands: Netherlands Organisation for Scientific Research (NWO Veni 2020 - VI.Veni.202.179); Norway: Research Council of Norway (RCN-314472); Poland: Ministry of Science and Higher Education (IDUB AGH, POB8, D4 no 9722), Polish National Agency for Academic Exchange (PPN/PPO/2020/1/00002/U/00001), Polish National Science Centre (NCN 2021/42/E/ST2/00350, NCN OPUS nr 2022/47/B/ST2/03059, NCN UMO-2019/34/E/ST2/00393, UMO-2020/37/B/ST2/01043, UMO-2021/40/C/ST2/00187, UMO-2022/47/O/ST2/00148, UMO-2023/49/B/ST2/04085); Slovenia: Slovenian Research Agency (ARIS grant J1-3010); Spain: Generalitat Valenciana (Artemisa, FEDER, IDIFEDER/2018/048), Ministry of Science and Innovation (MCIN \& NextGenEU PCI2022-135018-2, MICIN \& FEDER PID2021-125273NB, RYC2019-028510-I, RYC2020-030254-I, RYC2021-031273-I, RYC2022-038164-I), PROMETEO and GenT Programmes Generalitat Valenciana (CIDEGENT/2019/027); Sweden: Swedish Research Council (Swedish Research Council 2023-04654, VR 2018-00482, VR 2022-03845, VR 2022-04683, VR 2023-03403, VR grant 2021-03651), Knut and Alice Wallenberg Foundation (KAW 2018.0157, KAW 2018.0458, KAW 2019.0447, KAW 2022.0358); Switzerland: Swiss National Science Foundation (SNSF - PCEFP2\_194658); United Kingdom: Leverhulme Trust (Leverhulme Trust RPG-2020-004), Royal Society (NIF-R1-231091); United States of America: U.S. Department of Energy (ECA DE-AC02-76SF00515), Neubauer Family Foundation.

%
%


\clearpage
\printbibliography
\clearpage
 
\begin{flushleft}
\hypersetup{urlcolor=black}
{\Large The ATLAS Collaboration}

\bigskip

\AtlasOrcid[0000-0002-6665-4934]{G.~Aad}$^\textrm{\scriptsize 104}$,
\AtlasOrcid[0000-0001-7616-1554]{E.~Aakvaag}$^\textrm{\scriptsize 17}$,
\AtlasOrcid[0000-0002-5888-2734]{B.~Abbott}$^\textrm{\scriptsize 123}$,
\AtlasOrcid[0000-0002-0287-5869]{S.~Abdelhameed}$^\textrm{\scriptsize 119a}$,
\AtlasOrcid[0000-0002-1002-1652]{K.~Abeling}$^\textrm{\scriptsize 56}$,
\AtlasOrcid[0000-0001-5763-2760]{N.J.~Abicht}$^\textrm{\scriptsize 50}$,
\AtlasOrcid[0000-0002-8496-9294]{S.H.~Abidi}$^\textrm{\scriptsize 30}$,
\AtlasOrcid[0009-0003-6578-220X]{M.~Aboelela}$^\textrm{\scriptsize 45}$,
\AtlasOrcid[0000-0002-9987-2292]{A.~Aboulhorma}$^\textrm{\scriptsize 36e}$,
\AtlasOrcid[0000-0001-5329-6640]{H.~Abramowicz}$^\textrm{\scriptsize 154}$
% ...
\bigskip
\\

% ...
$^{17}$Department for Physics and Technology, University of Bergen, Bergen; Norway.\\
% ...
$^{30}$Physics Department, Brookhaven National Laboratory, Upton NY; United States of America.\\
% ...
$^{36}$$^{(a)}$Facult\'e des Sciences Ain Chock, Universit\'e Hassan II de Casablanca;$^{(b)}$Facult\'{e} des Sciences, Universit\'{e} Ibn-Tofail, K\'{e}nitra;$^{(c)}$Facult\'e des Sciences Semlalia, Universit\'e Cadi Ayyad, LPHEA-Marrakech;$^{(d)}$LPMR, Facult\'e des Sciences, Universit\'e Mohamed Premier, Oujda;$^{(e)}$Facult\'e des sciences, Universit\'e Mohammed V, Rabat;$^{(f)}$Institute of Applied Physics, Mohammed VI Polytechnic University, Ben Guerir; Morocco.\\
% ...
$^{45}$Physics Department, Southern Methodist University, Dallas TX; United States of America.\\
% ...
$^{50}$Fakult\"{a}t Physik , Technische Universit{\"a}t Dortmund, Dortmund; Germany.\\
% ...
$^{56}$II. Physikalisches Institut, Georg-August-Universit\"{a}t G\"ottingen, G\"ottingen; Germany.\\
% ...
$^{104}$CPPM, Aix-Marseille Universit\'e, CNRS/IN2P3, Marseille; France.\\
% ...
$^{119}$$^{(a)}$New York University Abu Dhabi, Abu Dhabi;$^{(b)}$United Arab Emirates University, Al Ain; United Arab Emirates.\\
% ...
$^{123}$Homer L. Dodge Department of Physics and Astronomy, University of Oklahoma, Norman OK; United States of America.\\
% ...
$^{154}$Raymond and Beverly Sackler School of Physics and Astronomy, Tel Aviv University, Tel Aviv; Israel.\\
% ...
$^{a}$ Also Affiliated with an institute covered by a cooperation agreement with CERN.\\
$^{b}$ Also at An-Najah National University, Nablus; Palestine.\\
$^{c}$ Also at Borough of Manhattan Community College, City University of New York, New York NY; United States of America.\\
$^{d}$ Also at Center for Interdisciplinary Research and Innovation (CIRI-AUTH), Thessaloniki; Greece.\\
$^{e}$ Also at Centro Studi e Ricerche Enrico Fermi; Italy.\\
$^{f}$ Also at CERN, Geneva; Switzerland.\\
% ...

\end{flushleft}


\end{document}